\documentclass[aps,prc,floatfix,twocolumn,nofootinbib,superscriptaddress]{revtex4-1}

\usepackage[]{graphicx}
\usepackage{color}

\usepackage{amsmath,amssymb,amsfonts}

\usepackage{slashed}

\begin{document}

\title{Medium-induced gluon emission via transverse and longitudinal scattering in dense nuclear matter}

\author{Le Zhang}

\affiliation{Institute of Particle Physics and Key Laboratory of Quark and Lepton Physics (MOE), Central China Normal University, Wuhan, 430079, China  }
\affiliation{The College of Post and Telecommunication, Wuhan Institute of Technology, Wuhan, 430070, China  }

\author{De-Fu Hou}
\affiliation{Institute of Particle Physics and Key Laboratory of Quark and Lepton Physics (MOE), Central China Normal University, Wuhan, 430079, China  }

\author{Guang-You Qin}
\affiliation{Institute of Particle Physics and Key Laboratory of Quark and Lepton Physics (MOE), Central China Normal University, Wuhan, 430079, China  }

\date{\today}
\begin{abstract}

We study the medium-induced gluon emission from a hard quark jet traversing the dense nuclear matter within the framework of deep inelastic scattering off a large nucleus.
We extend the previous work and compute the single gluon emission spectrum including both transverse and longitudinal momentum exchanges between the hard jet parton and the medium constituents.
On the other hand, with only transverse scattering and using static scattering centers for the traversed medium, our induced gluon emission spectrum in the soft gluon limit reduces to the Gyulassy-Levai-Vitev one-rescattering-one-emission formula.

\end{abstract}
\maketitle

\section{Introduction}

The study of parton energy loss and jet quenching has been regarded as a very useful tool to probe the properties of the quark-gluon plasma (QGP) produced in ultra-relativisitic heavy-ion collisions at the Relativistic Heavy-Ion Collider (RHIC) and the Large Hadron Collider (LHC) \cite{ Wang:1991xy, Qin:2015srf, Blaizot:2015lma, Majumder:2010qh}.
After they are produced from early stage hard collisions, high transverse momentum partonic jets propagate through the dense nuclear medium and interact with the medium constituents via binary elastic and radiative inelastic collisions before fragmenting into hadrons.
Jet-medium interaction not only changes the energy of the leading parton, but also modifies the internal structure of the jets, such as the distribution of jet momentum among different jet constituents.
The picture of jet-medium interaction and parton energy loss has been confirmed by many experimental observations at RHIC and the LHC, such as the suppression of high transverse momentum hadron \cite{Abelev:2012hxa, Aad:2015wga, CMS:2012aa} and jet \cite{Adam:2015ewa, Aad:2014bxa, Khachatryan:2016jfl} production, and the modification of jet correlations \cite{Aad:2010bu, Chatrchyan:2012gt} and jet structures \cite{Chatrchyan:2013kwa, Aad:2014wha}, in nucleus-nucleus collisions as compared to the expectations from independent nucleon-nucleon collisions.

Several theoretical schemes have been founded to study the radiative component of energy loss experienced by the hard jet partons propagating through the dense nuclear medium, such as Baier-Dokshitzer-Mueller-Peigne-Schiff-Zakharov (BDMPS-Z) \cite{Baier:1996kr, Baier:1996sk, Baier:1998kq, Zakharov:1996fv, Zakharov:1997uu}, Gyulassy, Levai and Vitev (GLV) \cite{Gyulassy:1999zd, Gyulassy:2000er},
Armesto-Salgado-Wiedemann (ASW) \cite{Wiedemann:2000za,Wiedemann:2000tf,Armesto:2003jh}, Arnold-Moore-Yaffe (AMY) \cite{Arnold:2001ba, Arnold:2002ja} and higher twist (HT) \cite{Guo:2000nz, Wang:2001ifa, Majumder:2009ge} formalisms.
For a more detailed comparison of different jet energy loss schemes, the reader is referred to Ref. \cite{Armesto:2011ht} and references therein.
There have also been many literatures studying the effect of binary elastic collisions between the hard partons and the medium constituents \cite{Bjorken:1982tu, Braaten:1991we, Djordjevic:2006tw, Qin:2007rn}.

While the studies of jet quenching in heavy-ion collisions have already entered the quantiative era, there still exist many theoretical uncertainties in detailed calculations of the effects caused by jet-medium interaction.
For example, most current studies of inelastic radiative contribution mainly focus on the gluon emission induced by the transverse momentum exchange between the propagating hard jet partons and the dense nuclear matter.
However, when a hard parton interacts with the traversed dense medium, both transverse and longitudinal momenta are exchanged between them \cite{Majumder:2008zg, Qin:2012fua, Abir:2014sxa}.
While there have been many studies on the longitudinal momentum exchange (loss) experienced by the propagating jet partons, the main focus was on the evaluation of purely collisional energy loss either from the leading parton \cite{Wicks:2005gt, Djordjevic:2006tw, Qin:2007rn, Schenke:2009ik, Cao:2013ita} or by the shower partons of the full jet \cite{Qin:2009uh, Neufeld:2009ep, Qin:2010mn, Qin:2012gp, Chang:2016gjp, Tachibana:2017syd}.
In Refs. \cite{Qin:2014mya, Zhang:2016avg, Abir:2015hta}, the effect of longitudinal momentum transfer between the hard parton and the medium constituents on the medium-induced emission vertex has been studied.

In this work, we study the medium-induced gluon emission from a hard quark jet which scatters off the medium constituents during the propagation through the dense nuclear medium, within the framework of deep-inelastic scattering (DIS).
We extend the HT radiative energy loss approach \cite{Guo:2000nz, Wang:2001ifa} and include the contributions from both transverse and longitudinal momentum exchanges to the gluon emission vertex.
It is also an extension of Refs. \cite{Qin:2014mya, Zhang:2016avg} which study the medium-induced photon emission from longitudinal and transverse scatterings.
Here we derive a closed formula for the medium-induced single gluon emission spectrum with the inclusion of the contributions from both transverse and longitudinal momentum transfers between the hard parton and the medium constituents.
We further show that if one neglects the longitudinal momentum transfer and only considers the transverse scattering, our medium-induced gluon emission spectrum in the soft gluon limit can reduce to the GLV one-rescattering-one-emission formula.

The paper is organized as follows. In Sec.~II, we present the gluon emission at leading twist in the DIS framework. In Sec.~III, we derive the medium-induced gluon emission spectrum for a quark jet parton traversing the dense nuclear medium. Some details and main results may be found in the Appendix. Sec.~IV contains our summary.

\section{Gluon emission at leading twist}

Here we study the gluon emission from a hard jet parton in dense nuclear matter in the framework of deep inelastic scattering (DIS) off a large nucleus.
We consider the following DIS process:
\begin{eqnarray}
 e(L_1) + A(P_A) \to e(L_2) + q(l_q) + g(l) + X,
\end{eqnarray}
where $L_1$ and $L_2$ are the momenta of the incoming and outgoing leptons, and $P_A = A p$ is the momentum of the incoming nucleus, with $p = [p^+, p^-, \mathbf{p}_\perp] = [p^+, 0, \mathbf{0}_\perp]$ being the momentum carried by each nucleon in the nucleus, $l_q$ and $l$ are the momenta of the produced hard quark and the radiated gluon.
Here the light-cone notation are used for four-vectors, e.g., $p^+ = (p^0+p^3)/\sqrt{2}$, and $p^- = (p^0 - p^3)/\sqrt{2}$.
In the Briet frame, the virtual photon $\gamma^*$ carries a momentum $q = L_2 - L_1 = [-x_B p^+, q^-, \mathbf{0}_\perp]$, with $x_B = Q^2 / (2p^+q^-)$ being the Bjorken fraction variable and $Q^2=-q^2$ the invariant mass of the virtual photon.

The differential cross section the lepton production from the above DIS process can be expressed as follows:
\begin{eqnarray}
E_{L_2} \frac{ d\sigma}{d^3\mathbf{L}_2} = \frac{\alpha^2_{e}}{2\pi s} \frac{1}{Q^4} L_{\mu\nu} W^{\mu\nu}.
\end{eqnarray}
Here $\alpha_e$ is the electromagnetic coupling and $s=(p+L_1)^2$ is the center-of-mass energy of the lepton-nucleon collision system.
The leptonic tensor $L_{\mu\nu}$ is given as:
\begin{eqnarray}
L_{\mu\nu} = \frac{1} {2} \rm Tr[\slashed{L}_1 \gamma_\mu  \slashed{L}_2 \gamma_\nu].
\end{eqnarray}
The hadronic tensor $W^{\mu\nu}$ is expressed as:
\begin{eqnarray}
W^{\mu\nu} &=&\sum_{X} (2\pi)^4 \delta^4 (q + P_A - P_X)
\nonumber\\  &\times &\langle A| J^\mu(0)|X\rangle \langle X|J^\nu(0)|A\rangle,
\end{eqnarray}
where $|A \rangle$ denotes initial state of the incoming nucleus A, and $|X \rangle$ represent the final hadronic (or partonic) states, with $\sum_X$ running over all possible final states except the outgoing hard quark jet and the emitted gluon.
$J^\mu = Q_q \bar{\psi} \gamma^\mu \psi $ is the hadron electromagnetic current for a quark of flavor $q$ and the electric charge $Q_q$ (in units of the electric charge $e$).
In our study, the focus is the hadronic tensor which contains the detailed information about the final state interaction between the struck hard quark and the traversed dense nuclear medium.

\begin{figure}[thb]
\includegraphics[width=0.99\linewidth]{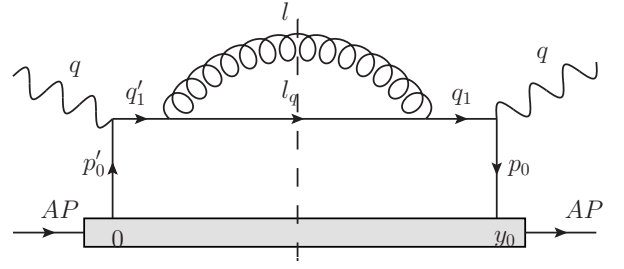}
 \caption{Leading twist contribution to the hadronic tensor.
} \label{gluon0}
\end{figure}

Figure~\ref{gluon0} shows the gluon emission process in semi-inclusive DIS at leading twist level. It represents the process that a hard quark is first excited by the virtual photon from one nucleon of the nucleus, then radiates a gluon and exits the medium without further interaction.
Here we use the light-cone gauge, $n\cdot A = A^-=0$, with $n=[1, 0, \mathbf{0}_\perp]$.
The sum of the gluon polarization in this light-cone gauge is given by:
\begin{eqnarray}
\tilde{G}^{\alpha \beta}(l)&=&-g^{\alpha \beta}+\frac{n^{\alpha}l^{\beta}+n^{\beta}l^{\alpha}}{n\cdot l}.
\end{eqnarray}
In the limit of very high energy and collinear emission, one may neglect the ($\perp$) component of the incoming quark field operators and factor out the one-nucleon state from nucleus state.
After some straightforward calculation, the contribution to the hadronic tensor at leading twist can be obtained as follows:
\begin{eqnarray}
\frac{dW_0^{A\mu\nu}}{dl_\perp^2 dy} &=& \sum_q Q_q^2  (-g_\perp^{\mu\nu})   A C_p^A  (2\pi)f_q(x_B+x_L) \nonumber\\
&\times&
\frac{\alpha_s}{2 \pi} C_F \frac{P(y)}{l_\perp^2}.
\end{eqnarray}
Here $g_\perp^{\mu\nu} = g^{\mu\nu} - g^{\mu-}g^{\nu+} - g^{\mu+} g^{\nu-}$, $C_p^A$ denotes the probability of finding a nucleon state with a momentum $p$ inside the nucleus $A$, $\alpha_s$ is the strong coupling, $y=l^-/q^-$ is the fraction of the forward momentum carried by the radiated gluon with respect to the parent quark, and $l_\perp$ is the gluon's transverse momentum. For convenience, the momentum fraction $x_L = l_\perp^2/[2p^+q^-y(1-y)]$ is also defined.
The quark parton distribution function $f_q(x)$ is defined as:
\begin{eqnarray}
f_q(x) = \int \frac{dy_0^-}{2\pi} e^{-ixp^+y_0^-} \langle p| \bar{\psi}(y_0^-) \frac{\gamma^+}{2} \psi(0)|p \rangle.
\end{eqnarray}
where $x$ is the fraction of the forward momentum carried by the quark from the nucleon.
The leading order quark-to-gluon (photon) splitting function $P(y)$ is given by:
\begin{eqnarray}
P(y)= \frac{1+(1-y)^2}{y}.
\end{eqnarray}
Note the color factor $C_F$ for quark to gluon splitting vertex is factored out.
Therefore, the differential single gluon emission spectrum at leading order (without rescattering with the medium constituents) is given as:
\begin{eqnarray}
\frac{dN_g^{\rm vac}}{dy dl_\perp^2} = \frac{\alpha_s}{2 \pi} C_F \frac{P(y)}{l_\perp^2} =
	\frac{\alpha_s}{2 \pi} C_F \frac{1+(1-y)^2}{y} \frac{1}{l_\perp^2}.
\end{eqnarray}

\section{Medium-induced gluon emission via single rescattering}

Now we compute the medium-induced gluon emission spectrum from a hard quark jet traversing the dense nuclear medium.
In this work, we use the following power counting scheme and notations: $Q$ for the hardest momentum scale, and $\lambda$ for a small dimensionless parameter.
Considering the scattering of a nearly on-shell projectile parton carrying a momentum $q = (q^+, q^-, {q}_\perp) \sim (\lambda^2 Q, Q, 0)$ off a nearly on-shell target parton carrying a momentum $\sim (Q, \lambda^2 Q, 0)$, the exchanged gluons is then the standard Glauber gluon carrying a momentum $\sim (\lambda^{2} Q, \lambda^{2} Q, \lambda Q)$ \cite{Idilbi:2008vm}.
If the target parton is allowed to be off shell, the longitudinal momentum component of the exchanged gluon may be the same order as the transverse components; such type of gluons is often referred to as the {longitudinal}-Glauber gluon which carries a momentum $\sim (\lambda^{2} Q, \lambda Q, \lambda Q)$ \cite{Qin:2012fua}.
In this work, we investigate the influence of both transverse and longitudinal momentum transfers on the single gluon emission from a hard quark jet which interacts with the constituents of the dense nuclear medium.

\begin{figure}[thb]
\includegraphics[width=0.99\linewidth]{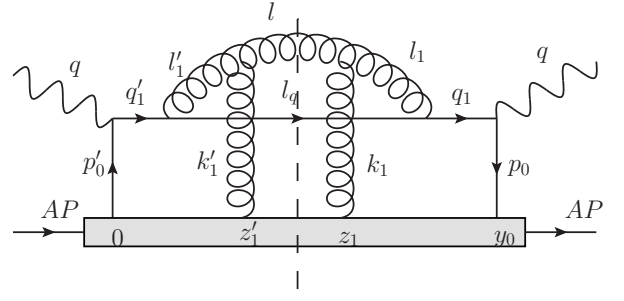}
 \caption{A central-cut diagram which contributes to the twist-4 hadronic tensor: one rescattering on the radiated gluon in both the amplitude and the complex conjugate.}
\label{gluon1}
\end{figure}

Figure~\ref{gluon1} shows one of central-cut diagrams that contributes to the hadronic tensor at the twist-four level.
It describes the process with a single rescattering on the radiated gluon in both the amplitude and the complex conjugate.
The other 20 diagrams at the twist-four level are presented in Appendix.
In this section, we provide the detailed calculation of the hadronic tensor and the medium-induced gluon emission spectrum for Figure~\ref{gluon1}.
The calculations for the other 20 diagrams are completely analogous and their main results are listed in Appendix.

In Figure~ \ref{gluon1}, a hard virtual photon carrying a momentum $q$ strikes a quark from the nucleus at the location $y_0'=0$ ($y_0$ in the complex conjugate).
The incoming quark from the nucleus carries a momentum $p_0'$ ($p_0$ in the complex conjugate).
The struck quark carries a momentum $q_1'$ ($q_1$ in the complex conjugate) and emits a gluon with a momentum $l'_1$ ($l_1$ in the complex conjugate).
The emitted gluon then scatters off the gluon field at the location $z_1'$ ($z_1$ in the complex conjugate) and picks up a momentum $k_1'$ ($k_1$ in the complex conjugate).
The final radiated gluon and the final outgoing quark carry the momenta $l$ and $l_q$, respectively.
The contribution to the hadronic tensor from Figure~ \ref{gluon1} can be written as follows:
\begin{widetext}
\begin{eqnarray}
 W^{A\mu\nu}_{(2)}&=&\sum_q Q_q^2\frac{1}{N_c} \int \frac{d^4 l}{{(2 \pi)}^4}2\pi \delta (l^2)\int \frac{d^4l_q}{{(2\pi)}^4}2\pi \delta (l_q^2)
 \int d^4 y_0 e^{i q y_0}\int d^4 z \int d^4 z' \int d^4 z_1 \int d^4 z_1'
 \nonumber\\ &\times&
 \int \frac{d^4 q_1}{{(2\pi)}^4}e^{-i q_1 \cdot (y_0-z)} \int \frac{d^4 l_1}{{(2\pi)}^4}e^{-i l_1 \cdot (z-z_1)} e^{-i l \cdot (z_1-z_1')} e^{-i l_q  \cdot (z-z')} \int \frac{d^4 l_1'}{{(2\pi)}^4}e^{-i l_1' \cdot (z_1'-z')} \int \frac{d^4 q_1'}{{(2\pi)}^4}e^{-i q_1' \cdot (z'-y_0')}
 \nonumber\\ &\times&
 \langle A | \bar{\psi}(y_0) \gamma^\mu \frac{-i \slashed{q}_1}{q_1^2-i \epsilon}(-i g \gamma_{\alpha_0} T^{a_0}) \slashed{l}_q G^{\alpha_0 \beta_1}_{a_0 b_1}(l_1) \Gamma_{\beta_1 \gamma_1 \alpha_1}^{b_1 c_1 a_1}(-l_1, -k_1,l) A_{c_1}^{\gamma_1}(z_1)\tilde{G}_{a_1 a'_1}^{\alpha_1 \alpha'_1}(l)
 \nonumber\\ &\times&
\Gamma_{\alpha'_1 \gamma'_1 \beta'_1}^{a'_1 c'_1 b'_1}(-l, k'_1,l'_1)  A_{c'_1}^{\gamma'_1}(z'_1) G^{\beta'_1 \alpha'_0 }_{b'_1 a'_0 }(l'_1) (i g\gamma_{\alpha'_0}T^{a'_0}) \frac{i\slashed{q}'_1}{q_1'^2+i \epsilon}\gamma^\nu \psi(0)|A \rangle.
\label{eq1}
\end{eqnarray}
Here, $\Gamma_{\alpha'_1 \gamma'_1 \beta'_1}^{a'_1 c'_1 b'_1}(-l, k'_1,l'_1)$ [$\Gamma_{\beta_1 \gamma_1 \alpha_1}^{b_1 c_1 a_1}(-l_1, -k_1,l) $  in complex conjugate] is the three-gluon vertex,
\begin{eqnarray}
\Gamma_{\beta_1 \gamma_1 \alpha_1}^{b_1 c_1 a_1}(-l_1, -k_1,l) &=& g f^{b_1 c_1 a_1} \tilde{\Gamma}_{\beta_1 \gamma_1 \alpha_1}(-l_1, -k_1,l),
\end{eqnarray}
where $f^{b_1 c_1 a_1}$ is the anti-symmetric structure constant of $SU(3)$ group, and
\begin{eqnarray}
\tilde{\Gamma}_{\beta_1 \gamma_1 \alpha_1}(-l_1, -k_1,l)
= g_{\beta_1 \gamma_1}{(-l_1+k_1)}_{\alpha_1} + g_{\gamma_1 \alpha_1}{(-k_1-l)}_{\beta_1} + g_{ \alpha_1 \beta_1}{(l+l_1)}_{\gamma_1} \,.
\end{eqnarray}
$G^{\beta'_1 \alpha'_0 }_{ b'_1  a'_0}(l'_1)$ [$G^{\alpha_0 \beta_1}_{a_0 b_1}(l_1)$  in complex conjugate] is the gluon propagator,
\begin{eqnarray}
G^{\alpha_0 \beta_1}_{a_0 b_1}(l_1) = \frac{-i}{l_1^2-i \epsilon} \tilde{G}^{\alpha_0 \beta_1}_{a_0 b_1}(l_1) = \frac{-i}{l_1^2-i \epsilon} \delta_{a_0 b_1} \tilde{G}^{\alpha_0 \beta_1}(l_1).
\end{eqnarray}

To simplify the hadronic tensor $W^{A\mu\nu}_{(2)}$, one first isolates the phase factors associated with the gluon insertions: $e^{-i(l_q + l_1 - q_1) \cdot z}  e^{i(l_q + l'_1 - q'_1) \cdot z'}$.
After carrying out the integration over $z$ and $z'$, we obtain two $\delta$ functions which may be used to integrate over the momenta $q_1$ and $q'_1$, rendering the relations: $q_1= l_q-l_1$ and $q'_1= l_q- l'_1$.
From the momentum conservation at each vertex, we can obtain the following relations among various momenta in Figure~\ref{gluon1}:
\begin{eqnarray}
p_0=q_1-q,\,\,\,\, k_1=l-l_1,\,\,\,\, p'_0=q'_1-q,\,\,\,\, k'_1=l-l'_1.
\end{eqnarray}
Re-introducing the momentum $p_0 = l+l_q - k_1 - q$ and changing the integration variables $l_1 \to k_1$ and $l'_1 \to k'_1$, the phase factor can be expressed as: $e^{-ip_0\cdot y_0} e^{-ik_1\cdot z_1} e^{ik'_1\cdot z'_1} $.

In this study, we work in the very high energy and collinear emission limit, and the dominant component of the rescattered gluon field is the forward $(+)$ component.
In such limit, one may factor out one-nucleon state from the nucleus state and ignore the $(\perp)$ component of the quark field operators,
\begin{eqnarray}
\langle A | \bar{\psi}(y_0) \gamma^\mu \hat{O} \gamma^\nu \psi(0) | A \rangle
\approx A C_p^A \langle p | \bar{\psi}(y_0) \frac{\gamma^+}{2} \psi(0) | p \rangle
 {\rm Tr} [\frac{\gamma^-}{2} \gamma^\mu \frac{\gamma^+}{2} \gamma^\nu]{\rm Tr} [\frac{\gamma^-}{2} \langle A| \hat{O} |A\rangle].
\end{eqnarray}
The gluon propagators together with the three-gluon vertices in the hard trace part may be simplified as:
\begin{eqnarray}
&&\tilde{G}^{\alpha_0 \beta_1}(l_1) \tilde{\Gamma}_{\beta_1 \gamma_1 \alpha_1}(-l_1, -k_1,l) A_{c_1}^{\gamma_1}(z_1) \tilde{G}^{\alpha_1 \alpha'_1}(l)\tilde{\Gamma}_{\alpha_1' \gamma_1' \beta'_1}(-l, k'_1,l'_1)A_{c'_1}^{\gamma'_1}(z'_1)\tilde{G}^{\beta'_1 \alpha'_0 }(l'_1)
\nonumber \\
&& = \tilde{G}^{\alpha_0 \beta_1}(l_1) \left[g_{\alpha_1 \beta_1} (l+l_1)^- A_{c_1}^+(z_1)\right]
\tilde{G}^{\alpha_1 \alpha'_1}(l) \left[g_{\alpha_1' \beta_1'} (l+l_1')^- A_{c_1'}^+(z_1') \right] \tilde{G}^{\beta'_1 \alpha'_0 }(l'_1)
\end{eqnarray}
With the above simplification, the hadronic tensor may be written as follows:
\begin{eqnarray}
 W^{A\mu\nu}_{(2)}&=&\sum_q Q_q^2 g^4 \int \frac{d^4 l}{{(2 \pi)}^4}2\pi \delta (l^2)\int \frac{d^4l_q}{{(2\pi)}^4}2\pi \delta (l_q^2)
 \int d^4 y_0  \int d^4 z_1 \int d^4 z'_1 \int \frac{d^4 k_1}{{(2\pi)}^4} \int \frac{d^4 k_1'}{{(2\pi)}^4}
\nonumber\\ 	&\times&
\int \frac{d^4p_0}{(2\pi)^4} (2\pi)^4 \delta^4(l+l_q - p_0 - k_1 - q) \left(e^{-ip_0\cdot y_0} e^{-ik_1\cdot z_1} e^{ik_1'\cdot z_1'}\right)
	 \frac{1}{q^2_1-i\epsilon}\frac{(l+l_1)^-}{l^2_1-i\epsilon}\frac{1}{q'^2_1+i\epsilon}\frac{(l +l_1')^-}{l'^2_1+i\epsilon}
 \nonumber\\ 	&\times&
(-g_\perp^{\mu\nu})A C_p^A \langle p | \bar{\psi}(y_0) \frac{\gamma^+}{2} \psi(0) | p \rangle \langle A | A_{c_1}^+ (z_1)  A_{c'_1}^+ (z'_1)| A \rangle \times \frac{1}{N_c}{\rm Tr}[T^{a_0}f^{a_0 c_1 a_1}f^{a_1 c'_1 a'_0} T^{a'_0}]
 \nonumber\\ 	&\times&
	{\rm Tr}\left[\frac{\gamma^-}{2} \slashed{q}_1 \gamma_{\alpha_0} \slashed{l}_q \gamma_{\alpha_0'}\slashed{q}_1'\right]
	\tilde{G}^{\alpha_0 \beta_1}(l_1) g_{\alpha_1 \beta_1} \tilde{G}^{\alpha_1 \alpha'_1}(l) g_{\alpha_1' \beta_1'} \tilde{G}^{\beta'_1 \alpha'_0 }(l'_1).
\label{eq3}
\end{eqnarray}

We now look at the internal propagators and external lines.
For the quark propagator before the gluon emission,
\begin{eqnarray}
q_1^2 = (q+p_0)^2 = 2p^+q^-(1+ x_0^-)[-x_B + x_0 - x_{D0}],
\end{eqnarray}
where we have defined the momentum factions,
\begin{eqnarray}
x_0 = \frac{p_0^+}{p^+},  \,  x_0^- = \frac{p_0^-}{q^-}, \, x_{D0}  = \frac{ \mathbf{p}_{0\perp}^2}{2p^+q^-(1+ x_0^-)}.
\end{eqnarray}
For the internal gluon propagator,
\begin{eqnarray}
l_1^2 = (l-k_1)^2 = 2p^+q^-(y- \lambda_1^-)[x_L(1-y) - \lambda_1 - \lambda_{D1}],
\end{eqnarray}
where we have defined the momentum factions,
\begin{eqnarray}
\lambda_1= \frac{k_1^+}{p^+}, \;\; \lambda_1^- = \frac{k_1^-}{q^-}, \;\;
\lambda_{D1} = \frac{(\mathbf{l}_{\perp}-\mathbf{k}_{1 \perp})^2}{2p^+q^-(y-\lambda_1^-)}.
\end{eqnarray}
For the final outgoing quark, the on-shell condition gives:
\begin{eqnarray}
(2\pi)\delta(l_q^2)=\frac{1}{2p^+q^-(1+ x_0^- + \lambda_1^- -y )}(2\pi)\delta(-x_B + x_0 +\lambda_1 - x_L(1-y) - \eta_{D1}).
\end{eqnarray}
where we have defined the momentum factions,
\begin{eqnarray}
\eta_{D1} = \frac{(\mathbf{l}_{\perp}-\mathbf{k}_{1 \perp}- \mathbf{p}_{0 \perp})^2}{2p^+q^-(1+ x_0^- + \lambda_1^- -y )}.
\end{eqnarray}
Combining the internal quark and gluon lines with the final outgoing quark,
\begin{eqnarray}
D_q &=&\frac{C_q }{(2q^-)^3 (p^+)^5}
\frac{1}{-x_B+x_0-x_{D0}-i\epsilon}  \frac{y- \frac{\lambda_1^-}{2}}{(y-\lambda_1^-)(x_L(1-y) - \lambda_1 - \lambda_{D1}-i\epsilon)}\frac{1}{-x_B+x_0' -x_{D0}' +i\epsilon}
\nonumber\\&\times&
 \frac{y - \frac{\lambda_1'^-}{2}}{(y-\lambda_1'^-)(x_L(1-y) - \lambda_1' - \lambda_{D1}'+i\epsilon)}
 (2\pi)\delta(-x_B + x_0 +\lambda_1 - x_L(1-y) - \eta_{D1}),
\end{eqnarray}
where
\begin{eqnarray}
C_q = \frac{1}{1+x_0^-} \frac{1}{1+ x_0'^-}\frac{1}{1+x_0^- + \lambda_1^- -y}.
\end{eqnarray}
The trace part of the hadronic tensor [the last line of Eq.~(\ref{eq3})] can be simplified as:
\begin{eqnarray}
\hspace{-5mm}
	N_q = \frac{4 q^-}{C_q}\frac{1+\left(1-\frac{y- \lambda_1^-}{1+ x_0^-}\right)\left(1-\frac{y- \lambda_1'^-}{1+ x_0'^-}\right)}{(y - \lambda_1^-)(y -\lambda_1'^-)\left(1-\frac{y- \lambda_1^-}{1+ x_0^-}\right)\left(1-\frac{y- \lambda_1'^-}{1+ x_0'^-}\right)}
	\left(\mathbf{l}_\perp-\mathbf{k}_{1 \perp}-\frac{y-\lambda_1^-}{1+ x_0^-}\mathbf{p}_{0 \perp}\right)\cdot\left(\mathbf{l}_\perp-\mathbf{k}_{1 \perp}'-\frac{y-\lambda_1'^-}{1+ x_0'^-}\mathbf{p}_{0 \perp}'\right).
\end{eqnarray}
With the above simplifications, the hadronic tensor now reads:
\begin{eqnarray}
 W^{A\mu\nu}_{(2)}&=&\sum_q Q_q^2 g^4 \int \frac{d^4 l}{{(2 \pi)}^4}2\pi \delta (l^2)\int \frac{d^4l_q}{{(2\pi)}^4}(2\pi)^4 \delta^4(l+l_q - p_0 - k_1 - q)
 \int d^4 y_0  \int d^4 z_1 \int d^4 z_1'
\nonumber\\ &\times&
 \int \frac{d^3 \mathbf{k}_1 d \lambda_1}{{(2\pi)}^4} \int \frac{d^3 \mathbf{k}_1' d \lambda_1'}{{(2\pi)}^4} \int \frac{d^3\mathbf{p}_0 d x_0}{(2\pi)^4} \left(e^{-ix_0 p^+ y_0^-} e^{-i\lambda_1 p^+  z_1^-} e^{i\lambda_1' p^+z_1'^-}\right)
 \left(e^{-i\mathbf{p}_0\cdot \mathbf{y}_0} e^{-i\mathbf{k}_1\cdot \mathbf{z}_1} e^{i\mathbf{k}_1'\cdot \mathbf{z}_1'}\right)
 \nonumber\\ &\times&(-g_\perp^{\mu\nu})A C_p^A \langle p | \bar{\psi}(y_0) \frac{\gamma^+}{2} \psi(0) | p \rangle \langle A | A_{c_1}^+ (z_1)  A_{c'_1}^+ (z'_1)| A \rangle \times \frac{1}{N_c}{\rm Tr}[T^{a_0}f^{a_0 c_1 a_1}f^{a_1 c'_1 a'_0} T^{a'_0}]
 \nonumber\\ &\times&
\frac{1}{-x_B+x_0-x_{D0}-i\epsilon}  \frac{1}{x_L(1-y) - \lambda_1 - \lambda_{D1}-i\epsilon}\frac{1}{-x_B+x_0' -x_{D0}' +i\epsilon}  \frac{1}{x_L(1-y) - \lambda_1' - \lambda_{D1}'+i\epsilon}
 \nonumber\\ &\times&
 (2\pi)\delta[-x_B + x_0 +\lambda_1 - x_L(1-y) - \eta_{D1}]
\frac{2}{{(2 p^+ q^-)}^2}\frac{1+\left(1-\frac{y- \lambda_1^-}{1+ x_0^-}\right)\left(1-\frac{y- \lambda_1'^-}{1+ x_0'^-}\right)}{(y - \lambda_1^-)(y - \lambda_1'^-)\left(1-\frac{y- \lambda_1^-}{1+ x_0^-}\right)\left(1-\frac{y- \lambda_1'^-}{1+ x_0'^-}\right)}
 \nonumber\\ &\times&
\left(\frac{y- \frac{\lambda_1^-}{2}}{y- \lambda_1^-}\right)\left(\frac{y - \frac{\lambda_1'^-}{2}}{y- \lambda_1'^-}\right)
 \left(\mathbf{l}_\perp - \mathbf{k}_{1 \perp}-\frac{y- \lambda_1^-}{1+ x_0^-}\mathbf{p}_{0 \perp} \right) \cdot \left(\mathbf{l}_\perp - \mathbf{k}_{1 \perp}'-\frac{y- \lambda_1'^-}{1+ x_0'^-}\mathbf{p}_{0 \perp}'\right),
\end{eqnarray}
where for convenience, we have used the three-vector notations for momentum and coordinate: $\mathbf{k} = (k^-, \mathbf{k}_{\perp})$ and $\mathbf{z} = (z^+, \mathbf{z}_{\perp})$, and $\mathbf{k} \cdot \mathbf{z} = k^-z^+ - \mathbf{k}_{\perp} \cdot \mathbf{z}_{\perp}$.

Now we perform the integration over the momentum fractions $x_0$, $\lambda_1$, and $\lambda_1'$. Using the on-shell condition for the outgoing quark and the overall momentum conservation $p'_0 = p_0 + k_1 - k'_1$, we can carry out the integration over $x_0$,
\begin{eqnarray}
&&\int \frac{dx_0}{2 \pi}\frac{e^{- i x_0 p^+ y_0^-}}{-x_B+x_0-x_{D0}-i\epsilon}\frac{1}{-x_B+{x}_0'-{x}_{D0}'+i\epsilon}(2\pi)\delta[-x_B + x_0 +\lambda_1 - x_L(1-y) - \eta_{D1}]
\nonumber\\
	&&= e^{- i (x_B +x_L(1-y) + \eta_{D1} ) p^+ y_0^-} \frac{e^{+ i \lambda_1 p^+ y_0^-}}{x_L(1-y) - \lambda_1 + \eta_{D1} - x_{D0}-i\epsilon}\frac{e^{-i \lambda_1' p^+ y_0'^-}}{x_L(1-y) - \lambda_1' + \eta_{D1}' - x_{D0}' + i\epsilon}.
\end{eqnarray}
The remaining integration over $\lambda_1$ may be performed with a counterclockwise semicircle in the upper half of the complex plane:
\begin{eqnarray}
&&\int \frac{d\lambda_1}{2 \pi}\frac{e^{- i \lambda_1 p^+ (z_1^--y_0^-)}}{(x_L(1-y) - \lambda_1 + \eta_{D1} - x_{D0}-i\epsilon)(x_L(1-y) - \lambda_1 - \lambda_{D1}-i\epsilon)}
\nonumber\\&&\,\, =\,\, i \theta(z_1^- -y_0^-)e^{-i x_L(1-y) p^+ (z_1^- - y_0^-)}e^{i (\eta_{D1} - x_{D0}) p^+ y_0^-} e^{i \lambda_{D1} p^+ z_1^-} \frac{e^{- i \chi_{D10} p^+ y_0^-} - e^{- i \chi_{D10} p^+ z_1^-} }{\chi_{D10}},
\end{eqnarray}
where we have defined the variable $\chi_{D10} = \eta_{D1} + \lambda_{D1} - x_{D0}$ for convenience.
The integration over the momentum fraction $\lambda_1'$ is completely analogous.
After performing the integration over the quark lines and the gluon lines, we obtain the hadronic tensor as:
\begin{eqnarray}
 W^{A\mu\nu}_{(2)}&=&\sum_q Q_q^2(-g_\perp^{\mu\nu})A C_p^A g^4 \int \frac{d^4 l}{{(2 \pi)}^4}2\pi \delta (l^2)\int \frac{d^4l_q}{{(2\pi)}^4}(2\pi)^4 \delta^4(l+l_q - p_0 - k_1 - q)
\\ &\times&
 \int d y_0^- \int d^3 \mathbf{y}_0 \int \frac{d^3\mathbf{p}_0}{(2\pi)^3} e^{-i\mathbf{p}_0\cdot \mathbf{y}_0}  e^{- i x_B p^+ y_0^-} \langle p | \bar{\psi}(y_0) \frac{\gamma^+}{2} \psi(0) | p \rangle
 \nonumber\\ &\times&
\int d z_1^- i \theta(z_1^- -y_0^-) \int d {z'_1}^- (-i) \theta(z_1'^- - y_0'^-) \int d^3 \mathbf{z}_1 \int \frac{d^3 \mathbf{k}_1 }{{(2\pi)}^3} \int d^3 \mathbf{z}_1'  \int \frac{d^3 \mathbf{k}_1' }{{(2\pi)}^3} \left( e^{-i\mathbf{k}_1\cdot \mathbf{z}_1} e^{i\mathbf{k}_1'\cdot \mathbf{z}_1'}\right)
 \nonumber\\ &\times&
 \langle A | A_{c_1}^+ (z_1)  A_{c_1'}^+ (z_1')| A \rangle \times \frac{1}{N_c}{\rm Tr}[T^{a_0}f^{a_0 c_1 a_1}f^{a_1 c_1' a_0'} T^{a_0'}]
  \nonumber\\ &\times&
 e^{- i x_{D0} p^+ y_0^-} e^{- i (x_L(1-y)-\lambda_{D1}) p^+ z_1^-} e^{ i (x_L(1-y)-\lambda_{D1}') p^+ z_1'^-} (e^{- i \chi_{D10} p^+ y_0^-} - e^{- i \chi_{D10} p^+ z_1^-})(1 - e^{i \chi_{D10}' p^+ z_1'^-})
 \nonumber\\ &\times&
\frac{1}{\chi_{D10}}\frac{1}{\chi_{D10}'}\frac{2}{{(2 p^+ q^-)}^2}\frac{1+\left(1-\frac{y- \lambda_1^-}{1+ x_0^-}\right)\left(1-\frac{y- \lambda_1'^-}{1+ x_0'^-}\right)}{(y - \lambda_1^-)(y - \lambda_1'^-)\left(1-\frac{y-\lambda_1^-}{1+ x_0^-}\right)\left(1-\frac{y- \lambda_1'^-}{1+ x_0'^-}\right)}
\nonumber \\&\times& \left(\frac{y- \frac{\lambda_1^-}{2}}{y- \lambda_1^-}\right)\left(\frac{y - \frac{\lambda_1'^-}{2}}{y- \lambda_1'^-}\right)\left(\mathbf{l}_\perp - \mathbf{k}_{1 \perp}-\frac{y- \lambda_1^-}{1+ x_0^-}\mathbf{p}_{0 \perp} \right)\cdot \left(\mathbf{l}_\perp - \mathbf{k}_{1 \perp}'-\frac{y- \lambda_1'^-}{1+ x_0'^-}\mathbf{p}_{0 \perp}'\right). \nonumber
\end{eqnarray}

To proceed, one may write the expectation of two gluon operators in the nucleus state as follows:
\begin{eqnarray}
 \langle A |  A_{c_1}^+(z_1) A_{c_1'}^+(z_1')  |A\rangle
	= \frac{1}{d(R)} {\rm Tr} [T_{c_1}(R) T_{c_1'}(R)] \langle A |  A^+(z_1) A^+(z_1')  |A\rangle
	= \frac{\delta_{c_1 c_1'} C_2(R)}{N_c^2-1} \langle A| A^+(z_1) A^+(z_1') | A\rangle,
\end{eqnarray}
where $d(R)$ and $C_2(R)$ are the dimension and the Casimir factor of the representation $R$, with $C_2(R) = C_F$ and $C_2(R) = C_A$ for exchanging the gluon field initiated by the on-shell quark and gluon, respectively.
Note that two quark operators have been factored out, and two gluon insertions have the same color.
Then the color factor for the hadronic tensor may be evaluated as:
\begin{eqnarray}
	\frac{\delta_{c_1 c_1'}}{N_c} {\rm Tr}[T^{a_0}f^{a_0 c_1 a_1}f^{a_1 c_1' a_0'} T^{a_0'}] = C_F C_A.
\end{eqnarray}
Now we make transformation for the coordinate variables $(z_1, z_1')\rightarrow (Z_1, \delta z_1)$:
$Z_1 = (z_1 + z_1') /2, \delta z_1 = z_1 - z_1'$.
The translational invariance of the correlation functions gives:
\begin{eqnarray}
\label{translational invariance}
\langle A| A^+(z_1) A^+(z'_1) | A\rangle \approx \langle A| A^+(\delta{z_1}) A^+(0) | A\rangle.
\end{eqnarray}
With the above transformation, the integration for the phase factor may now be carried out,
\begin{eqnarray}
\int d^3\mathbf{z}_1 \int d^3 \mathbf{z}'_1 e^{-i\mathbf{k}_1 \cdot \mathbf{z}_1} e^{i\mathbf{k}'_1\cdot \mathbf{z}'_1}
= (2\pi)^3 \delta^3(\mathbf{k}_1 - \mathbf{k}'_1) \int d^3 \mathbf{\delta z}_1 e^{-i\mathbf{k}_1\cdot \delta\mathbf{z}_1}.
\end{eqnarray}
The above $\delta$ function implies that the  pair of gluon field insertions in each nucleon state carry the same momentum, $\mathbf{k}'_1 = \mathbf{k}_1$.
In addition, one may carry out the integration over the space coordinate $\mathbf{y}_0$ and obtain $\mathbf{p}_0=\mathbf{p}'_0=0$. Then the hadronic tensor may be written as:
\begin{eqnarray}
	W^{A\mu\nu}_{(2)} &=&\sum_q Q_q^2(-g_\perp^{\mu\nu}) A C_p^A \frac{C_2(R)}{N_c^2-1} \frac{\alpha_s^2}{\pi}
	\int\frac{dy}{y}  \int d^2 \mathbf{l}_\perp
	\int d y_0^- e^{- i (x_B+x_L) p^+ y_0^-} \langle p | \bar{\psi}(y_0) \frac{\gamma^+}{2} \psi(0) | p \rangle
	\\ &\times&
\int d Z_1^- \int d\delta z_1^-  \int d^3 \mathbf{\delta z}_1 \int \frac{d^3\mathbf{k}_1}{(2\pi)^3} e^{-i\mathbf{k}_1\cdot \delta \mathbf{z}_1} \langle A | A^+(\delta z_1^-, \delta \mathbf{z}_1) A^+(0)  |A\rangle
\nonumber \\ &\times&
e^{i x_L p^+ y_0^-}  e^{-i x_{D0} p^+ y_0^-} e^{- i (x_L(1-y)-\lambda_{D1}) p^+ \delta z_1^-}(e^{- i \chi_{D10} p^+ y_0^-} - e^{- i \chi_{D10} p^+ (Z_1^- +\frac{1}{2} \delta z_1^-)})(1 - e^{i \chi_{D10} p^+ (Z_1^- -\frac{1}{2} \delta z_1^-)})
 \nonumber\\ &\times&
C_A C_F \frac{1}{{(\chi_{D10})}^2}\frac{2}{{(2 p^+ q^-)}^2}\frac{1+{\left(1 + \lambda_1^- - y\right)}^2}{{(y - \lambda_1^-)}^2{\left(1 + \lambda_1^- -y \right)}^2}
{\left(\frac{y-\frac{\lambda_1^-}{2}}{y- \lambda_1^-}\right)}^2 {\left(\mathbf{l}_\perp - \mathbf{k}_{1 \perp} \right)}^2 . \nonumber
\end{eqnarray}

Now we simplify the phase factor (the second last line in the above hadronic tensor, denoted as $S_{(2)}$).
We first note that $Z_1^-$ is the location of the gluon insertion point which can span over the nucleus size, while $y_0^-$ and $\delta z_1^-$ are confined within the nucleon size, thus $y_0^-, \delta z_1^- \ll Z_1^-$. This may be used to simplify the phase factor:
\begin{eqnarray}
\label{deltas2}
S_{(2)} &=&  e^{i x_L p^+ y_0^-} e^{- i x_{D0} p^+ y_0^-} e^{- i (x_L(1-y)-\lambda_{D1}) p^+ \delta z_1^-}(e^{- i \chi_{D10} p^+ y_0^-} - e^{- i \chi_{D10} p^+ (Z_1^- +\frac{1}{2} \delta z_1^-)})(1 - e^{i \chi_{D10} p^+ (Z_1^- -\frac{1}{2} \delta z_1^-)})
\nonumber\\  &\approx& [2-2 \cos(\chi_{D10} p^+ Z_1^-)].
\end{eqnarray}
Now we recall the expression of $\chi_{D10}$:
\begin{eqnarray}
\chi_{D10} = \eta_{D1} + \lambda_{D1} - x_{D0} =\frac{{\left(\mathbf{l}_\perp - \mathbf{k}_{1 \perp} \right)}^2}{2p^+ q^-(y - \lambda_1^-)\left(1+\lambda_1^--y\right)}
	= x_L \frac{y (1-y)}{(y - \lambda_1^-)(1+\lambda_1^--y)} \frac{{\left(\mathbf{l}_\perp - \mathbf{k}_{1 \perp} \right)}^2}{l_\perp^2}.
\end{eqnarray}
The hard matrix element (the last line of the above hadronic tensor, denoted as $\delta T_{(2)}$) may be simplified as:
\begin{eqnarray}
\label{deltat2}
	\delta T_{(2)} &=& \frac{2 y P(y)}{l_\perp^2} C_A C_F \left[\frac{1+(1+ \lambda_1^- -y)^2}{1+(1-y)^2} {\left(\frac{y-\frac{\lambda_1^-}{2}}{y- \lambda_1^-}\right)}^2  \frac{l_\perp^2}{{\left(\mathbf{l}_\perp - \mathbf{k}_{1 \perp} \right)}^2} \right] = \frac{2 y P(y)}{l_\perp^2} \delta \bar{T}_{(2)},
\end{eqnarray}
where we have defined $\delta \bar{T}_{(2)}$  for convenience.
With the above simplification, the hadronic tensor reads:
\begin{eqnarray}
W^{A\mu\nu}_{(2)} &=&\sum_q Q_q^2(-g_\perp^{\mu\nu}) A C_p^A (2\pi)f_q(x_B+x_L)
	\frac{C_2(R)}{N_c^2-1} (2\alpha_s^2)
	\int dy P(y) \int \frac{d^2 \mathbf{l}_\perp}{\pi l_\perp^2}
 \\ &\times&
\int d Z_1^-
\int d\delta z_1^-
	\int d^3 \mathbf{\delta z}_1 \int \frac{d^3\mathbf{k}_1}{(2\pi)^3} e^{-i\mathbf{k}_1\cdot \delta \mathbf{z}_1} \langle A | A^+(\delta z_1^-, \delta \mathbf{z}_1) A^+(0)  |A\rangle
	S_{(2)}(y, \mathbf{l}_{\perp}, \mathbf{k}_{\perp}, Z_1^-)
	\delta\bar{T}_{(2)}(y, \mathbf{l}_{\perp}, \mathbf{k}_{\perp})
.\nonumber
\end{eqnarray}
The above formula is quite general in the sense that the property of the dense nuclear medium that the hard jet parton probes is contained in the gluon field correlator $\langle A | A^+(\delta z_1) A^+(0)  |A\rangle$.
As long as the gluon field correlator (in momentum space) is known, one can use it to study the medium modification effect on the gluon emission process.
To perform the Fourier transformation for the gluon field, we come back to the Cartesian coordinate $z = (z^0, \mathbf{z}) = (z^0, z^1, z^2, z^3)$:
\begin{eqnarray}
\langle A | A^\mu(z_1) A^\nu(z_1')| A \rangle =
	\langle A | A^\mu(z_1^0, \mathbf{z}_1) A^\nu({z_1'}^0, \mathbf{z}_1')| A \rangle = \langle A |A^\mu(\mathbf{z}_1) A^\nu(\mathbf{z}_1')| A \rangle,
\end{eqnarray}
where for simplicity we have taken the gluon fields to be time-independent.
Then one may transform the gluon field correlation function into the momentum space as follows:
\begin{eqnarray}
\langle A |A^\mu(\mathbf{z}_1) A^\nu(\mathbf{z}_1')| A \rangle
	&=& \int \frac{d^3 \mathbf{p}}{(2 \pi)^3}e^{i \mathbf{p} \cdot \mathbf{z}_1} A^\mu(\mathbf{p})\int \frac{d^3 \mathbf{p}'}{(2 \pi)^3}e^{i \mathbf{p}' \cdot \mathbf{z}_1'} A^\nu(\mathbf{p}')
\nonumber\\
	&=& \int \frac{d^3 \bar{\mathbf{p}}}{(2 \pi)^3}\int \frac{d^3 \delta \mathbf{p}}{(2 \pi)^3}e^{i \delta \mathbf{p} \cdot \mathbf{Z}_1}e^{i \bar{\mathbf{p}} \cdot {\delta \mathbf{z}_1}} A^\mu(\bar{\mathbf{p}} + \delta \mathbf{p}/2)A^\nu(\bar{\mathbf{p}} - \delta \mathbf{p}/2).
\end{eqnarray}
Here, we have changed the integral variables from $d^3 \mathbf{p} d^3\mathbf{p}'$ to $d^3 \bar{\mathbf{p}} d^3 \delta \mathbf{p}$.
Using the translational invariance, the above gluon field correlator is not dependent on $\mathbf{Z}_1$, which implies $\delta \mathbf{p}=0$. Therefore, the gluon field correlation function may be written as:
\begin{eqnarray}
	\langle A | A^\mu(\delta \mathbf{z}_1) A^\nu(0)| A \rangle &=& \rho \int \frac{d^3 {\mathbf{p}}}{(2 \pi)^3}e^{i {\mathbf{p}} \cdot {\delta \mathbf{z}_1}} A^\mu({\mathbf{p}} )A^\nu({\mathbf{p}} )
	= \rho \int \frac{dp^3 d^2 {\mathbf{p}_\perp}}{(2 \pi)^3}e^{i ({p^3 \delta z_1^3 + \mathbf{p}_\perp} \cdot {\delta \mathbf{z}_{1\perp}})} A^\mu(p^3, {\mathbf{p}_\perp} )A^\nu(p^3, {\mathbf{p}_\perp} ),
\end{eqnarray}
where $\rho$ is the density of the medium constituents (scattering centers) that the hard jet parton interacts.
Now we substitute the above gluon field correlator and obtain the hadronic tensor as follows:
\begin{eqnarray}
W^{A\mu\nu}_{(2)} &=&\sum_q Q_q^2(-g_\perp^{\mu\nu}) A C_p^A (2\pi)f_q(x_B+x_L)
	\frac{C_2(R)}{N_c^2-1} (2\alpha_s^2)
	\int dy P(y) \int \frac{d^2 \mathbf{l}_\perp}{\pi l_\perp^2}
 \\ &\times&
\int d Z_1^-
\int d\delta z_1^-
	\int \frac{\sqrt{2}dk_1^- d^2\mathbf{k}_1}{(2\pi)^3} \rho A^+(\sqrt{2}k_1^-, {\mathbf{k}_{1\perp}} )A^+(\sqrt{2}k_1^-, {\mathbf{k}_{1\perp}} )
	S_{(2)}(y, \mathbf{l}_{\perp}, \lambda_1^-, \mathbf{k}_{1\perp}, Z_1^-)
	\delta\bar{T}_{(2)}(y, \mathbf{l}_{\perp}, \lambda_1^-, \mathbf{k}_{1\perp})
.\nonumber
\end{eqnarray}
Now we have to specify the form for the gluon field. For simplicity we take the static Yukawa potential:
\begin{eqnarray}
	A^{0}(\mathbf{p})=\frac{g}{\mathbf{p}^2+\mu^2}, \;\; \mathbf{A}(\mathbf{p})= 0.
\end{eqnarray}
where $\mu$ is the mass of the exchanged gluon. Note that the color factor for the gluon field has been factored out.
The gluon field correlation function now becomes:
\begin{eqnarray}
	\langle A | A^\mu(\delta z^0, \delta \mathbf{z}) A^\nu(0)| A \rangle = \delta^\mu_0 \delta^\nu_0 \rho \int \frac{d^3 {\mathbf{p}}}{(2 \pi)^3}e^{i {\mathbf{p}} \cdot {\delta \mathbf{z}}}\frac{g^2}{(\mathbf{p}^2 + \mu^2)^2}.
\end{eqnarray}
Then the hadronic tensor for Figure~\ref{gluon1} takes the following expression:
\begin{eqnarray}
	W^{A\mu\nu}_{(2)} &=&\sum_q Q_q^2(-g_\perp^{\mu\nu}) A C_p^A (2\pi)f_q(x_B+x_L) \frac{C_2(R)}{N_c^2-1} C_A C_F (2\alpha_s^2)
	\int dy P(y) \int \frac{d^2 \mathbf{l}_\perp}{\pi l_\perp^2}
 \\ &\times&
\int d Z_1^- \frac{\rho r_N^-}{2}
	\int \frac{\sqrt{2}dk_1^- d^2\mathbf{k}_{1\perp}}{(2\pi)^3}
	\frac{4\pi\alpha_s}{\left[(\sqrt{2}k_1^-)^2 + \mathbf{k}_{1\perp}^2 +\mu^2\right]^2}
\nonumber
 \\ &\times&
	\left[2-2 \cos\left( \frac{y (1-y)}{(y - \lambda_1^-)(1+\lambda_1^--y)} \frac{{\left(\mathbf{l}_\perp - \mathbf{k}_{1 \perp} \right)}^2}{l_\perp^2} \frac{Z_1^-}{\tau_{\rm form}^-} \right)\right]
		\left[\frac{1+(1+ \lambda_1^- -y)^2}{1+(1-y)^2}
	{\left(\frac{y-\frac{\lambda_1^-}{2}}{y- \lambda_1^-}\right)}^2  \frac{l_\perp^2}{{\left(\mathbf{l}_\perp - \mathbf{k}_{1 \perp} \right)}^2} \right],
\nonumber
\end{eqnarray}
where $r_N^- = \int d\delta z_1^-$ and $\tau_{\rm form}^- = 1/(x_L p^+) = 2q^-y(1-y)/l_\perp^2 = 2l^-(1-y)/l_\perp^2$ is the formation time of the radiated gluon.
The calculations for the other $20$ diagrams are completely analogous; their main results are provided in Appendix. After combining the contributions of all $21$ diagrams, the medium-induced single gluon emission spectrum reads:
\begin{eqnarray}
	\frac{d {N_g^{\rm med}}}{d y d^2 \mathbf{l}_{\perp}} &=& \frac{\alpha_s}{2\pi} \frac{C_2(R)}{N_c^2-1} {C_F} \frac{P(y)}{\pi l^2_{\perp}} \int d Z_1^-  (8 \sqrt{2} \pi^2 \alpha_s^2 \rho r_N^- )
	\int \frac{d k_1^-}{2\pi} \int \frac{d^2\mathbf{k}_{1\perp}}{(2\pi)^2}
	\frac{1}{\left[(\sqrt{2}k_1^-)^2 + \mathbf{k}_{1\perp}^2 +\mu^2\right]^2}
 \nonumber\\ &\times&
\left\{C_A\left[2-2 \cos\left(\frac{y(1-y)}{(y-\lambda_1^-)(1+\lambda_1^- -y)}\frac{{\left(\mathbf{l}_{\perp} - \mathbf{k}_{1 \perp} \right)}^2}{l_{\perp}^2} \frac{Z_1^-}{\tau_{\rm form}^-} \right)\right]
\nonumber\right.\\ & &\times\left[\frac{1+(1+ \lambda_1^- -y)^2}{1+(1-y)^2}
 {\left(\frac{y-\frac{\lambda_1^-}{2}}{y- \lambda_1^-}\right)}^2 \frac{l_{\perp}^2}{{\left(\mathbf{l}_{\perp} - \mathbf{k}_{1 \perp} \right)}^2}
	- \frac{1}{2} \frac{1+\left(1+\lambda_1^- - y \right)\left(1-y\right)}{1+(1-y)^2}\left( \frac{y-\frac{\lambda_1^-}{2}}{y-\lambda_1^-}\right)\frac{\mathbf{l}_{\perp} \cdot \left(\mathbf{l}_{\perp} - \mathbf{k}_{1 \perp}\right)}{\left(\mathbf{l}_{\perp} - \mathbf{k}_{1 \perp}\right)^2}
	\nonumber \right.\\ & & \left. - \frac{1}{2} \frac{1+\left(1+\lambda_1^- - y \right)\left(1-\frac{y}{1+\lambda_1^-}\right)}{1+(1-y)^2} \left(\frac{y-\frac{\lambda_1^-}{2}}{y-\lambda_1^-}\right)\frac{l_{\perp}^2\left(\mathbf{l}_{\perp} -\mathbf{k}_{1 \perp} \right)\cdot \left(\mathbf{l}_{\perp} - \frac{y}{1 + \lambda_1^-}\mathbf{k}_{1 \perp}\right)}{\left(\mathbf{l}_{\perp} -\mathbf{k}_{1 \perp} \right)^2 \left(\mathbf{l}_{\perp} - \frac{y}{1 + \lambda_1^-}\mathbf{k}_{1 \perp}\right)^2}
 \right]\nonumber\\ & &
	+\frac{C_A}{2}\left[2 - 2\cos \left( \frac{Z_1^-}{\tau_{\rm form}^-} \right)\right]\left[\frac{1+\left(1 - y \right)\left(1-\frac{y}{1+ \lambda_1^-}\right)}{1+(1-y)^2}\frac{\mathbf{l}_{\perp} \cdot \left(\mathbf{l}_{\perp} - \frac{y}{1+ \lambda_1^-} \mathbf{k}_{1\perp}\right)}{ \left(\mathbf{l}_{\perp} - \frac{y}{1+ \lambda_1^-} \mathbf{k}_{1\perp}\right)^2}-\frac{\left(y - \frac{\lambda_1^-}{2}\right)^2}{y (y - \lambda_1^-)} \right ]
 \nonumber\\ & & + C_F\left[2 - 2\cos \left( \frac{Z_1^-}{\tau_{\rm form}^-} \right)\right]\left[1 - \frac{1+\left(1 - y \right)\left(1-\frac{y}{1+ \lambda_1^-}\right)}{1+(1-y)^2}\frac{\mathbf{l}_{\perp} \cdot \left(\mathbf{l}_{\perp} - \frac{y}{1+ \lambda_1^-} \mathbf{k}_{1\perp}\right)}{  \left(\mathbf{l}_{\perp} - \frac{y}{1+ \lambda_1^-} \mathbf{k}_{1\perp}\right)^2}\right]
 \nonumber\\ & & \left.+C_F\left[\frac{1+\left(1-\frac{y}{1+ \lambda_1^-}\right)^2}{1+(1-y)^2}\frac{l_{\perp} ^2}{\left(\mathbf{l}_{\perp}  - \frac{y}{1+ \lambda_1^-} \mathbf{k}_{1\perp}\right)^2} -1\right]\right\}.
	\label{eq:dNgdydlt2_TL}
\end{eqnarray}
The above formula is one main result of the paper; it includes the contributions from both transverse momentum exchange ($\mathbf{k}_{1\perp}$) and longitudinal momentum transfer ($k_1^- = \lambda_1^- q^-$) to medium-induced gluon emission.

Now we concentrate on the medium-induced gluon emission induced only by transverse momentum exchange and neglect the contribution from longitudinal momentum transfer.
We will show that in the soft gluon emission limit, our result with only transverse scattering reduces to the GLV one-rescattering-one-emission formula \cite{Gyulassy:2000er, Gyulassy:2000gk, Djordjevic:2008iz, Buzzatti:2011vt}.
We first note that with only transverse momentum exchange between jet parton and medium, the gluon correlator reads:
\begin{eqnarray}
	\langle A | A^\mu(\delta \mathbf{z}_1) A^\nu(0)| A \rangle &=& \rho \int \frac{d^3 {\mathbf{p}}}{(2 \pi)^3}e^{i {\mathbf{p}} \cdot {\delta \mathbf{z}_1}} A^\mu({\mathbf{p}} )A^\nu({\mathbf{p}} )
	= \rho \delta(\delta z_1^3) \int \frac{d^2 {\mathbf{p}_\perp}}{(2 \pi)^2}e^{i \mathbf{p}_\perp \cdot \delta \mathbf{z}_{1\perp}} A^\mu({\mathbf{p}_\perp} )A^\nu({\mathbf{p}_\perp} ),
\end{eqnarray}
Using the static Yukawa potential, the medium-induced single gluon emission spectrum (after summing over all 21 diagrams) reads:
\begin{eqnarray}
	\frac{d {N_g^{\rm med}}}{d y d^2\mathbf{l}_{\perp}}&=& \frac{\alpha_s}{2\pi} \frac{C_2(R)}{N_c^2-1} {C_F} \frac{P(y)}{\pi l^2_{\perp}}  \int dZ_1^- (8 \sqrt{2}\pi^2 \alpha_s^2 \rho ) \int \frac{d^2\mathbf{k}_{1\perp}}{(2\pi)^2}\frac{1}{(\mathbf{k}_{1 \perp}^2 + \mu^2)^2}
 \nonumber\\ &\times&
\left\{C_A \left[2-2 \cos\left(\frac{{\left(\mathbf{l}_{\perp} - \mathbf{k}_{1 \perp} \right)}^2}{l_{\perp}^2} \frac{Z_1^-}{\tau_{\rm form}^-} \right)\right]
	\left[\frac{l_{\perp}^2}{{\left(\mathbf{l}_{\perp} - \mathbf{k}_{1 \perp} \right)}^2}
	- \frac{1}{2} \frac{\mathbf{l}_{\perp} \cdot \left(\mathbf{l}_{\perp} - \mathbf{k}_{1 \perp}\right)}{\left(\mathbf{l}_{\perp} - \mathbf{k}_{1 \perp}\right)^2}
\right.
	- \frac{1}{2} \frac{l_{\perp}^2\left(\mathbf{l}_{\perp} -\mathbf{k}_{1 \perp} \right)\cdot \left(\mathbf{l}_{\perp} - {y}\mathbf{k}_{1 \perp}\right)}{\left(\mathbf{l}_{\perp} -\mathbf{k}_{1 \perp} \right)^2 \left(\mathbf{l}_{\perp} - {y}\mathbf{k}_{1 \perp}\right)^2}\right]
\nonumber \\ & &
	+ \left(\frac{C_A}{2} - C_F \right) \left[2 - 2\cos \left( \frac{Z_1^-}{\tau_{\rm form}^-} \right)\right]\left[\frac{\mathbf{l}_{\perp} \cdot \left(\mathbf{l}_{\perp} - {y} \mathbf{k}_{1\perp}\right)}{ \left(\mathbf{l}_{\perp} - {y} \mathbf{k}_{1\perp}\right)^2} - 1\right ]
\left.+C_F\left[\frac{l_{\perp} ^2}{\left(\mathbf{l}_{\perp}  - {y}\mathbf{k}_{1\perp}\right)^2} -1\right]\right\}.
\end{eqnarray}
We note that the differential elastic cross section for a quark jet scattering with the Yukawa potential is:
\begin{eqnarray}
	\frac{d\sigma_{\rm el}}{d^2 \mathbf{p}_\perp} =  C_F \frac{ C_2(R) }{N_c^2-1}\frac{ |g A^0(\mathbf{p}_{\perp}) |^2}{4\pi^2} =  C_F \frac{C_2(R)}{N_c^2-1} \frac{4\alpha_s^2}{(\mathbf{p}_\perp^2 + \mu^2)^2}.
\end{eqnarray}
In very high energy limit, the total elastic cross section is obtained as:
\begin{eqnarray}
	\sigma_{\rm el} = C_F \frac{C_2(R)}{N_c^2-1} \frac{4\pi\alpha_s^2}{\mu^2}.
\end{eqnarray}
Therefore, one may write the elastic scattering probability as follows:
\begin{eqnarray}
\frac{dP_{\rm el}}{d^2 \mathbf{p}_\perp} = \frac{1}{\sigma_{\rm el}} \frac{d\sigma_{\rm el}}{d^2 \mathbf{p}_\perp} = \frac{\mu^2}{\pi(\mathbf{p}_\perp^2 + \mu^2)^2}.
\end{eqnarray}
Now note that $\rho\sigma_{\rm el} = {1}/{\lambda_{\rm mfp}} =  {\sqrt{2}/\lambda_{\rm mfp}^-}$. Therefore, we obtain the single gluon emission spectrum induced by transverse scattering as follows:
\begin{eqnarray}
	\frac{d {N_g^{\rm med}}}{d y d^2 \mathbf{l}_{\perp}}&=& \frac{\alpha_s}{2\pi} \frac{P(y)}{\pi l^2_{\perp}}  \int \frac{dZ_1^-}{\lambda_{\rm mfp}^-} \int {d^2\mathbf{k}_{1\perp}}
	\frac{1}{\sigma_{\rm el}} \frac{d\sigma_{\rm el}}{d^2 \mathbf{k}_{1\perp}}
 \nonumber\\ &\times&
\left\{C_A\left[2-2 \cos\left(\frac{{\left(\mathbf{l}_{\perp} - \mathbf{k}_{1 \perp} \right)}^2}{l_{\perp}^2}  \frac{Z_1^-}{\tau_{\rm form}^-} \right)\right]
	\left[\frac{l_{\perp}^2}{{\left(\mathbf{l}_{\perp} - \mathbf{k}_{1 \perp} \right)}^2}
	- \frac{1}{2} \frac{\mathbf{l}_{\perp} \cdot \left(\mathbf{l}_{\perp} - \mathbf{k}_{1 \perp}\right)}{\left(\mathbf{l}_{\perp} - \mathbf{k}_{1 \perp}\right)^2}
\right.
	- \frac{1}{2} \frac{l_{\perp}^2\left(\mathbf{l}_{\perp} -\mathbf{k}_{1 \perp} \right)\cdot \left(\mathbf{l}_{\perp} - {y}\mathbf{k}_{1 \perp}\right)}{\left(\mathbf{l}_{\perp} -\mathbf{k}_{1 \perp} \right)^2 \left(\mathbf{l}_{\perp} - {y}\mathbf{k}_{1 \perp}\right)^2}\right]
\nonumber \\ & &
+ \left(\frac{C_A}{2} - C_F\right) \left[2 - 2\cos \left(  \frac{Z_1^-}{\tau_{\rm form}^-} \right)\right]\left[\frac{\mathbf{l}_{\perp} \cdot \left(\mathbf{l}_{\perp} - {y} \mathbf{k}_{1\perp}\right)}{\left(\mathbf{l}_{\perp} - {y} \mathbf{k}_{1\perp}\right)^2}-1 \right ]
\left.+C_F\left[\frac{l_{\perp} ^2}{\left(\mathbf{l}_{\perp}  - {y}\mathbf{k}_{1\perp}\right)^2} -1\right]\right\}.
\end{eqnarray}
The above formula is the most important consequence of the calculations presented in this paper, which represents the medium-induced single gluon emission spectrum when considering only the transverse momentum exchange between the hard jet parton and the dense nuclear medium.
The transport property of the dense nuclear medium that the jet parton interacts is encoded in the differential elastic scattering rate $\frac{d\Gamma_{\rm el}}{d^2\mathbf{k}_{1\perp}dZ_1^-} 
= \frac{1}{\lambda_{\rm mfp}^-} \frac{1}{\sigma_{\rm el}} \frac{d\sigma_{\rm el}}{d^2\mathbf{k}_{1\perp}}$, which depends on the density of the medium constituents $\rho$ and the differential elastic scattering cross section ${d\sigma_{\rm el}}/{d^2\mathbf{k}_{1\perp}}$.
We note that although the above result only considers the contribution from the transverse scattering and
neglects the longitudinal momentum transfer, there is no assumption for the radiated gluon.
If we further takes the soft gluon emission limit ($y = l^-/q^- \ll 1$), the medium-induced single gluon emission spectrum becomes:
\begin{eqnarray}
	\frac{d {N_g^{\rm med}}}{d y d^2\mathbf{l}_\perp} &=&
	\frac{\alpha_s}{2\pi} \frac{P(y)}{\pi l_\perp^2} \int \frac{dZ_1^-}{\lambda_{\rm mfp}^-} \int {d^2\mathbf{k}_{1\perp}}
	\frac{1}{\sigma_{\rm el}} \frac{d\sigma_{\rm el}}{d^2 \mathbf{k}_{1\perp}}
C_A \left[2-2 \cos\left(\frac{{\left(\mathbf{l}_{\perp} - \mathbf{k}_{1 \perp} \right)}^2}{l_{\perp}^2}  \frac{Z_1^-}{\tau_{\rm form}^-} \right)\right]
\left[\frac{\mathbf{l}_{\perp} \cdot \mathbf{k}_{1 \perp}}{\left(\mathbf{l}_{\perp} - \mathbf{k}_{1 \perp}\right)^2}\right].
\end{eqnarray}
Note that in the small $y$ limit, the splitting function $P(y)$ reduces to $2/y$ and the formation time $\tau_{\rm form}^-$ reduces to $2q^-y/l_\perp^2 = 2l^-/l_\perp^2$ in the above expression.
Now our result agrees with the GLV one-rescattering-one-emission formula~\cite{Gyulassy:2000er, Gyulassy:2000gk, Djordjevic:2008iz, Buzzatti:2011vt}, see e.g., Eq. (5) in Ref. \cite{Gyulassy:2000gk} and  Eq. (1) in Ref. \cite{Djordjevic:2008iz} in static case.
Therefore, we recover the GLV result when considering only the transverse momentum exchange and taking the limit of soft gluon emission.

\end{widetext}

\section{Summary}

In this work, we have studied the medium-induced gluon emission from a hard quark jet traversing the dense nuclear medium within the framework of deep-inelastic scattering off a large nucleus.
In particular, we have extended the higher twist radiative enregy loss approach and computed the medium-induced single gluon emission spectrum including both transverse and longitudinal momentum transfers between the hard jet parton and the constituents of the nuclear medium.
We have also shown that our medium-induced gluon emission spectrum in the soft gluon limit can reduce to the Gyulassy-Levai-Vitev one-rescattering-one-emission formula if one considers only the contribution from transverse momentum exchange and assumes static scattering centers for the traversed medium.
Our study constitutes a significant progress in understanding the medium-induced radiative process during the interaction of the hard jet with the dense nuclear matter.
Phenomenological studies for parton energy loss and jet quenching in relativistic heavy-ion collisions will be presented in future publication.

\section*{ACKNOWLEDGMENTS}

We thank S. Cao, X.-N. Wang and Y. Zhang for discussions. This work is supported in part by Natural Science Foundation of China (NSFC) under grant Nos. 11775095, 11375072. D.-F.H. is supported by Ministry of Science and Technology of China (MSTC) under ``973" project No. 2015CB856904(4) and by NSFC under grant Nos. 11735007, 11375070, 11521064.

\begin{widetext}
\section*{APPENDIX}

In this Appendix, we present the main results (kernels) for the other 20 cut diagrams, as shown in Figures~(\ref{heavy2}-\ref{heavy12}).

\begin{figure}[thb]
\centering
\includegraphics[width=0.99\linewidth]{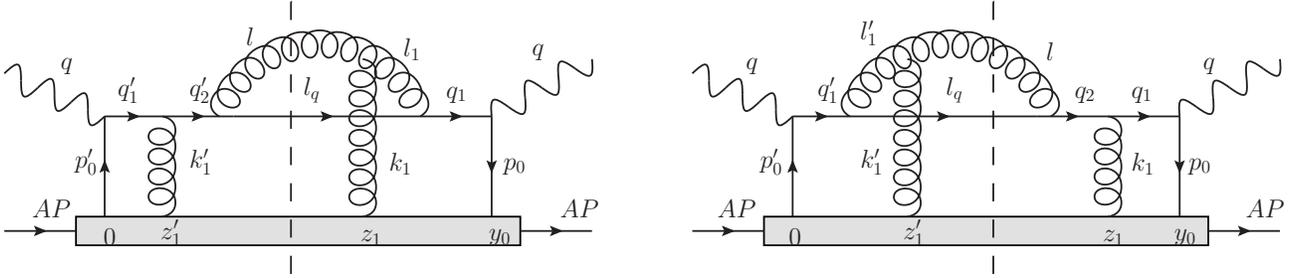}
 \caption{Two central-cut diagrams: one rescattering in both the amplitude and the complex conjugate.} 
 \label{heavy2}
\end{figure}

The phase factor for Figure~\ref{heavy2} reads:
\begin{eqnarray}
&&S_{(\ref{heavy2}),L}\,\,=\,\,e^{ix_L p^+ y_0^-} e^{- i x_{D0} p^+ y_0^-} e^{- i (x_L(1-y)-\frac{\lambda_{D1}}{2}-\frac{\eta_{D1}}{2}) p^+ \delta z_1^-}\left( e^{i\chi_{D10} p^+ Z_1^-} -1\right),
\nonumber\\
&&S_{(\ref{heavy2}),R}\,\,=\,\,e^{ix_L p^+ y_0^-} e^{i (x_L(1-y)-\frac{\lambda_{D1}}{2}-\frac{\eta_{D1}}{2}) p^+ \delta z_1^-}\left( e^{- i\chi_{D10} p^+ Z_1^-} -1\right),
\nonumber\\
&&S_{(\ref{heavy2})}\,\,=\,\,S_{(\ref{heavy2}),L} + S_{(\ref{heavy2}),R} \approx 2\cos(\chi_{D10} p^+ Z_1^-)-2.
\end{eqnarray}
The matrix element for Figure~\ref{heavy2} reads:
\begin{eqnarray}
& & \delta \bar{T}_{(\ref{heavy2}),L} =\delta \bar{T}_{(\ref{heavy2}),R}
= C_F\frac{C_A}{2}\left[\frac{1+\left(1+\lambda_1^- - y \right)\left(1-\frac{y}{1+\lambda_1^-}\right)}{1+(1-y)^2}\left( \frac{y-\frac{\lambda_1^-}{2}}{y-\lambda_1^-}\right)\frac{l_{ \perp}^2\left(\mathbf{l}_{\perp} -\mathbf{k}_{1 \perp} \right)\cdot \left(\mathbf{l}_{\perp} - \frac{y}{1 + \lambda_1^-}\mathbf{k}_{1 \perp}\right)}{\left(\mathbf{l}_{\perp} -\mathbf{k}_{1 \perp} \right)^2 \left(\mathbf{l}_{\perp} - \frac{y}{1 + \lambda_1^-}\mathbf{k}_{1 \perp}\right)^2} \right ].
\end{eqnarray}

\begin{figure}[thb]
\centering
\includegraphics[width=0.99\linewidth]{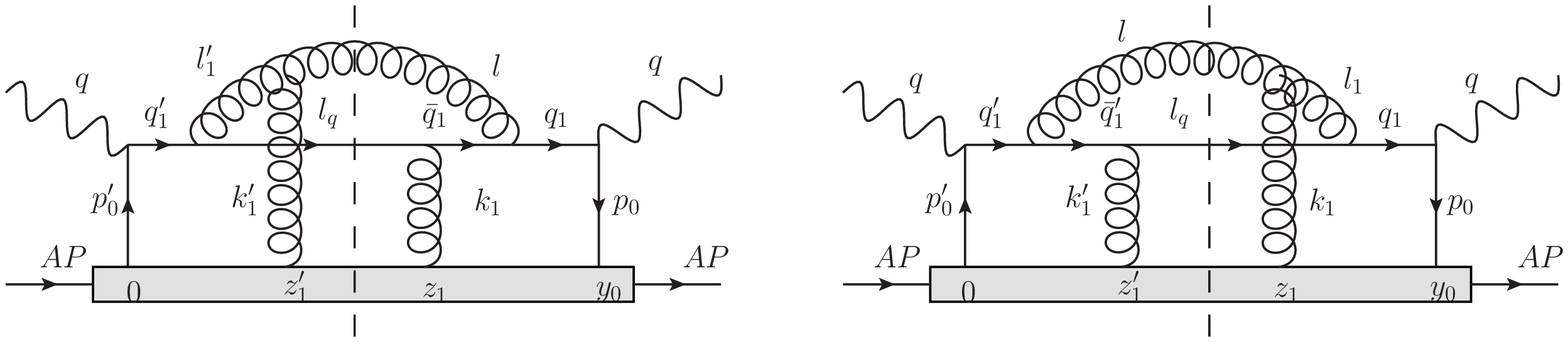}
 \caption{Two central-cut diagrams: one rescattering in both the amplitude and the complex conjugate.}
 \label{heavy3}
\end{figure}

The phase factor for Figure~\ref{heavy3} reads:
\begin{eqnarray}
&&S_{(\ref{heavy3}),L}\,\,=\,\, e^{ix_L p^+ y_0^-} e^{- i x_{D0} p^+ y_0^-}e^{- i (x_L(1-y)+\eta_{D1}) p^+  \delta z_1^-}(1 - e^{i x_L p^+ Z_1^-} - e^{-i \chi_{D10} p^+ Z_1^-} + e^{-i (\chi_{D10}-x_L) p^+ Z_1^-}),
\nonumber\\
&&S_{(\ref{heavy3}),R}\,\,=\,\, e^{ix_L p^+ y_0^-}e^{i (x_L(1-y)+\eta_{D1}) p^+  \delta z_1^-}(1 - e^{-i x_L p^+ Z_1^-} - e^{+i \chi_{D10} p^+ Z_1^-} + e^{i (\chi_{D10}-x_L) p^+ Z_1^-}),
\nonumber\\
&&S_{(\ref{heavy3})}\,\,=\,\,S_{(\ref{heavy3}),L}+S_{(\ref{heavy3}),R}\,\,\approx\,\,2 -2\cos( x_L p^+ Z_1^-) - 2\cos( \chi_{D10} p^+ Z_1^-) + 2\cos[(\chi_{D10}-x_L) p^+ Z_1^-].
\end{eqnarray}
The matrix element for Figure~\ref{heavy3} reads:
\begin{eqnarray}
\delta \bar{T}_{(4),L} =\delta \bar{T}_{(\ref{heavy3}),R}
&=& - C_F\frac{C_A}{2}\left[\frac{1+\left(1+\lambda_1^- - y \right)\left(1-y\right)}{1+(1-y)^2}\left( \frac{y-\frac{\lambda_1^-}{2}}{y-\lambda_1^-}\right)\frac{\mathbf{l}_{\perp} \cdot \left(\mathbf{l}_{\perp}- \mathbf{k}_{1 \perp}\right)}{\left(\mathbf{l}_{\perp}- \mathbf{k}_{1 \perp}\right)^2} \right].
\end{eqnarray}

\begin{figure}[thb]
\centering
\includegraphics[width=0.49\linewidth]{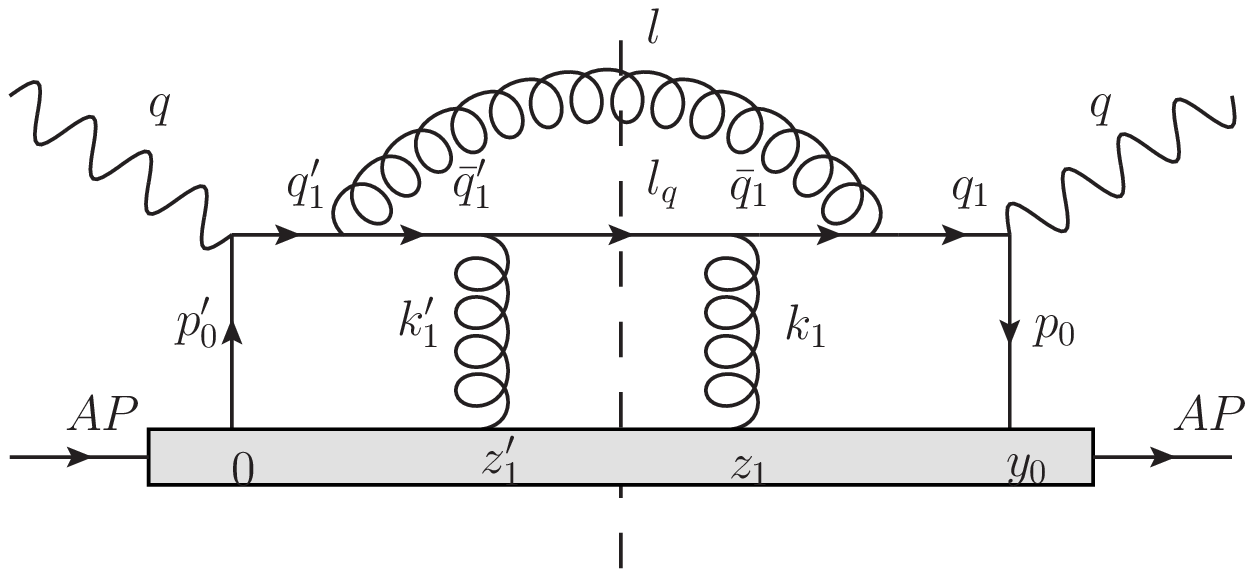}
 \caption{A central-cut diagram: one rescattering in both the amplitude and the complex conjugate.}
 \label{heavy4}
\end{figure}

The phase factor for Figure~\ref{heavy4} reads:
\begin{eqnarray}
S_{(\ref{heavy4})}&=& e^{ix_Lp^+y_0^-}  e^{-ix_{D0}p^+y_0^-} e^{- i (\eta_{D1}-\eta_{D0}) p^+  \delta z_1^-}(e^{-ix_Lp^+y_0^-}-e^{-ix_Lp^+Z_1^-})(1 - e^{ix_Lp^+Z_1^-})
\nonumber\\ &\approx& 2-2 \cos(x_L p^+ Z_1^-),
\end{eqnarray}
where
\begin{eqnarray}
\eta_{D0} = \frac{l_{\perp}^2}{2p^+q^-(1 - y )}.
\end{eqnarray}
The matrix element for Figure~\ref{heavy4} reads:
\begin{eqnarray}
\delta \bar{T}_{(\ref{heavy4})}
&=& C_F^2 .
\end{eqnarray}

\begin{figure}[thb]
\centering
\includegraphics[width=0.49\linewidth]{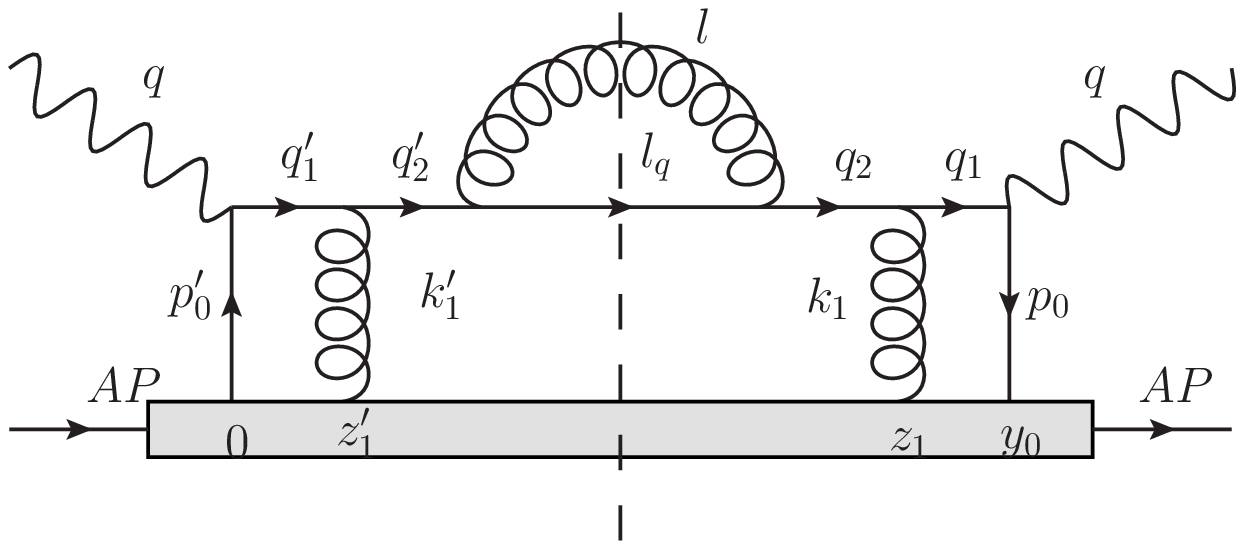}
\caption{A central-cut diagram: one rescattering in both the amplitude and the complex conjugate.}
  \label{heavy5}
\end{figure}

The phase factor for Figure~\ref{heavy5} reads:
\begin{eqnarray}
S_{(\ref{heavy5})}&=& e^{ix_Lp^+y_0^-} e^{-ix_{D0}p^+y_0^-} e^{- i x_{D1} p^+ \delta z_1^-} \approx 1,
\end{eqnarray}
where
\begin{eqnarray}
x_{D1} = \frac{k_{1 \perp}^2}{2p^+q^-(1+ \lambda_1^-)}.
\end{eqnarray}
The matrix element for Figure~\ref{heavy5} reads:
\begin{eqnarray}
\delta \bar{T}_{(\ref{heavy5})}
&=& C_F^2\left[\frac{1+\left(1-\frac{y}{1+ \lambda_1^-}\right)^2}{1+(1-y)^2}\frac{l_{ \perp}^2}{\left(\mathbf{l}_{\perp} - \frac{y}{1+ \lambda_1^-} \mathbf{k}_{1\perp}\right)^2} \right ].
\end{eqnarray}

\begin{figure}[thb]
\centering
\includegraphics[width=0.99\linewidth]{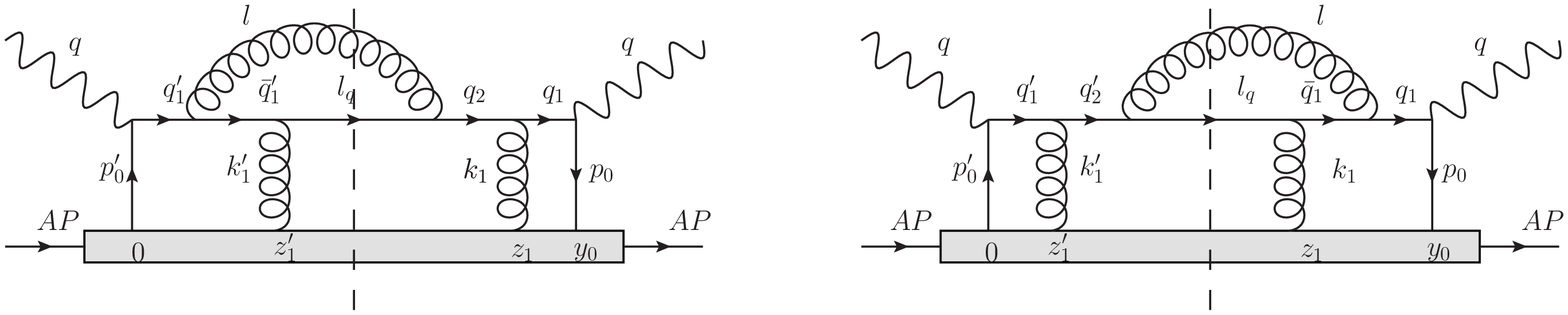}
\caption{Two central-cut diagrams: one rescattering in both the amplitude and the complex conjugate.}
  \label{heavy6}
\end{figure}

The phase factor for Figure~\ref{heavy6} reads:
\begin{eqnarray}
&&S_{(\ref{heavy6}),L}\,\,=\,\,e^{i x_L p^+ y_0^-}e^{- i ( x_L (1-y)+\eta_{D1}) p^+ \delta z_1^-}(e^{i x_L p^+ Z_1^-}-1),
\nonumber\\
&&S_{(\ref{heavy6}),R}\,\,=\,\,e^{i x_L p^+ y_0^-}e^{i ( x_L (1-y)+\eta_{D1}) p^+ \delta z_1^-}(e^{-i x_L p^+ Z_1^-}-1),
\nonumber\\
&&S_{(\ref{heavy6})}\,\,=\,\,S_{(\ref{heavy6}),L}\,\,+\,\,S_{(\ref{heavy6}),R}\,\,\approx\,\,2 \cos(x_L p^+ Z_1^-) -2.
\end{eqnarray}
The matrix element for Figure~\ref{heavy6} reads:
\begin{eqnarray}
&&\delta \bar{T}_{(\ref{heavy6}),L}=\delta \bar{T}_{(\ref{heavy6}),R}
= C_F \left( C_F-\frac{C_A}{2}\right)\left[\frac{1+\left(1 - y \right)\left(1-\frac{y}{1+ \lambda_1^-}\right)}{1+(1-y)^2} \frac{\mathbf{l}_{\perp} \cdot \left(\mathbf{l}_{\perp} - \frac{y}{1+ \lambda_1^-} \mathbf{k}_{1\perp}\right)}{ \left(\mathbf{l}_{\perp} - \frac{y}{1+ \lambda_1^-} \mathbf{k}_{1\perp}\right)^2} \right ].
\end{eqnarray}

\begin{figure}[thb]
\centering
\includegraphics[width=0.99\linewidth]{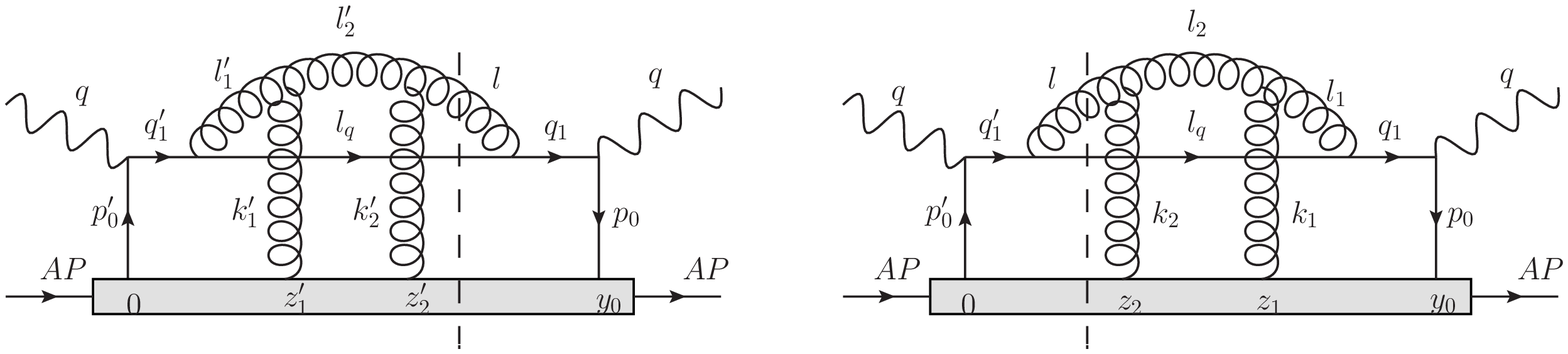}
 \caption{Two non-central-cut diagrams: two rescatterings in the amplitude and zero in the complex conjugate (or vice versa).}
 \label{heavy7}
\end{figure}

The phase factor for Figure~\ref{heavy7} reads:
\begin{eqnarray}
&& S_{(\ref{heavy7}),L}\,\,=\,\,\frac{1}{2}e^{- i (x_L(1-y)-\lambda_{D2}) p^+ \delta z_1^-}(e^{i x_L p^+ Z_1^-}-1),
\nonumber\\
&& S_{(\ref{heavy7}),R}\,\,=\,\,\frac{1}{2}e^{-i x_{D0} p^+ y_0^-}e^{i (x_L(1-y)-\lambda_{D2}) p^+ \delta z_1^-}(e^{-i x_L p^+ Z_1^-}-1),
\nonumber\\
&& S_{(\ref{heavy7})}\,\,=\,\,S_{(\ref{heavy7}),L}\,\,+\,\,S_{(\ref{heavy7}),R}\,\,\approx\,\, \cos(x_L p^+ Z_1^-) -1.
\end{eqnarray}
Here,
\begin{eqnarray}
\lambda_{D2} = \frac{(\mathbf{l}_{\perp}-\mathbf{k}_{2 \perp})^2}{2p^+q^-(y - \lambda_2^-)}.
\end{eqnarray}
The matrix element for Figure~\ref{heavy7} reads:
\begin{eqnarray}
 \delta \bar{T}_{(\ref{heavy7}),L}=\delta \bar{T}_{(\ref{heavy7}),R}
 & =& C_F C_A \frac{\left(y -\frac{\lambda_1^-}{2}\right)^2}{y ( y - \lambda_1^-)}.
\end{eqnarray}

\begin{figure}[thb]
\centering
\includegraphics[width=0.99\linewidth]{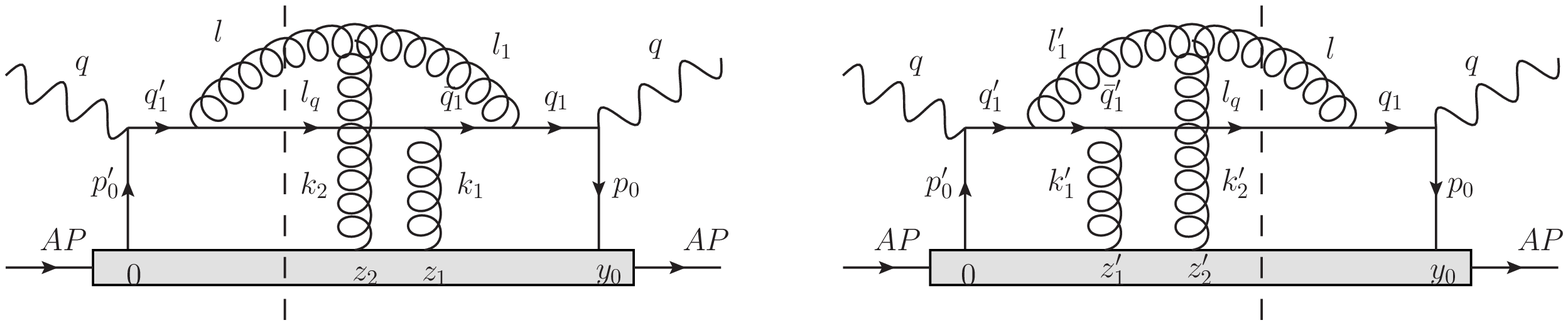}
 \caption{Two non-central-cut diagrams: two rescatterings in the amplitude and zero in the complex conjugate (or vice versa).}
 \label{heavy8}
\end{figure}

The phase factor for Figure~\ref{heavy8} reads:
\begin{eqnarray}
&& S_{(\ref{heavy8}),L}\,\,=\,\,e^{- i \frac{1}{2}(x_L(1-y)-\lambda_{D2}+\eta_{D2}-\bar{\eta}_{D2}) p^+ \delta z_1^-}( e^{-i x_L p^+ Z_1^- }- e^{-i(x_L-\chi_{D20}) p^+ Z_1^-} ),
\nonumber\\
&& S_{(\ref{heavy8}),R}\,\,=\,\,e^{-i x_{D0} p^+ y_0^-}e^{i \frac{1}{2}(x_L(1-y)-\lambda_{D2}+\eta_{D2}-\bar{\eta}_{D2}) p^+ \delta z_1^-}(e^{i x_L p^+ Z_1^- }- e^{i(x_L-\chi_{D20}) p^+ Z_1^-} ),
\nonumber\\
&& S_{(\ref{heavy8})}\,\,=\,\,S_{(\ref{heavy8}),L}\,\,+\,\,S_{(\ref{heavy8}),R}\,\,\approx\,\,2 \cos(x_L p^+ Z_1^-) - 2\cos[(x_L-\chi_{D20}) p^+ Z_1^-],
\end{eqnarray}
where
\begin{eqnarray}
\eta_{D2} &=& \frac{(\mathbf{l}_{ \perp} - \mathbf{k}_{1 \perp})^2}{2p^+q^-(1+\lambda_1^- -y )},\,\,\,\bar{\eta}_{D2} = \frac{\mathbf{l}_{ \perp}^2}{2p^+q^-(1 - y )}=y x_L,\\
\chi_{D20}&=&\eta_{D2}+ \lambda_{D2}-x_{D0}=\frac{{\left(\mathbf{l}_{ \perp} - \mathbf{k}_{1 \perp} \right)}^2}{2p^+ q^-(y - \lambda_1^-)\left(1+\lambda_1^--y\right)}= \chi_{D10}.
\end{eqnarray}
Note that we do not specify the time order for two interaction vertices $z_1$ and $z_2$ in Figure~\ref{heavy8} (Left) [$z_1'$ and $z_2'$ in Figure~\ref{heavy8} (Right)].
This explains the omission of the two diagrams (displayed in other papers e.g. \cite{Wang:2001ifa, Abir:2015hta}): one corresponding to the right cut of the left diagram in Figure~\ref{heavy8} and one corresponding to the left cut of the right diagram in Figure~\ref{heavy8}, which are already contained in the right and left diagrams in Figure~\ref{heavy8}, respectively.
The matrix element for Figure~\ref{heavy8} reads:
\begin{eqnarray}
& & \delta \bar{T}_{(\ref{heavy8}),L}=\delta \bar{T}_{(\ref{heavy8}),R}
=- C_F \frac{C_A}{2}\left[\frac{1+\left(1+\lambda_1^- - y \right)\left(1-y\right)}{1+(1-y)^2}\left(\frac{y - \frac{\lambda_1^-}{2}}{y - \lambda_1^-}\right)\frac{ \mathbf{l}_{ \perp} \cdot \left(\mathbf{l}_{ \perp} - \mathbf{k}_{1 \perp}\right)}{\left(\mathbf{l}_{ \perp} -\mathbf{k}_{1 \perp} \right)^2 } \right ].\nonumber
\end{eqnarray}

\begin{figure}[thb]
\centering
\includegraphics[width=0.99\linewidth]{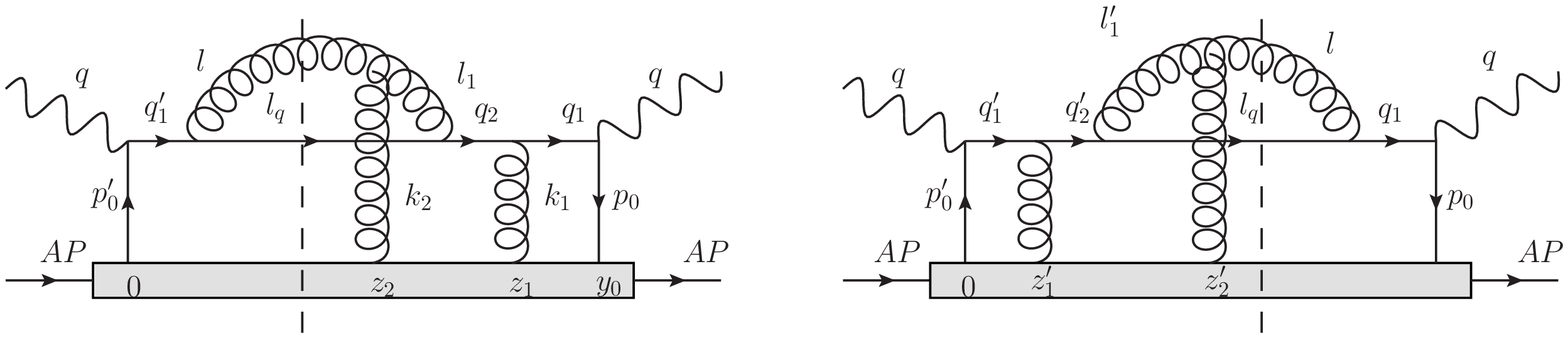}
 \caption{Two non-central-cut diagrams: two rescatterings in the amplitude and zero in the complex conjugate (or vice versa).}
 \label{heavy9}
\end{figure}

The phase factor for Figure~\ref{heavy9} reads:
\begin{eqnarray}
&& S_{(\ref{heavy9}),L}\,\,=\,\,\frac{1}{2} e^{ix_L p^+y_0^-}e^{- i x_L p^+ Z_1^-}\left(e^{i (x_L - 2\bar{\eta}_{D2} - 2 \lambda_{D2}) p^+ \frac{1}{2}\delta z_1^-} - e^{i (x_L  - 2 x_{D1}) p^+ \frac{1}{2}\delta z_1^-}\right),
\nonumber\\
&& S_{(\ref{heavy9}),R}\,\,=\,\,\frac{1}{2} e^{ix_L p^+y_0^-} e^{-ix_{D0}p^+y_0^-}e^{i x_L p^+ Z_1^-}\left(e^{-i (x_L - 2\bar{\eta}_{D2} - 2 \lambda_{D2}) p^+ \frac{1}{2}\delta z_1^-} - e^{-i (x_L - 2 x_{D1}) p^+ \frac{1}{2}\delta z_1^-}\right),
\nonumber\\
&& S_{(\ref{heavy9})}\,\,=\,\,S_{(\ref{heavy9}),L}\,\,+\,\,S_{(\ref{heavy9}),R}\,\,\approx\,\, 0.
\end{eqnarray}
The matrix element for Figure~\ref{heavy9} reads:
\begin{eqnarray}
& & \delta \bar{T}_{(\ref{heavy9}),L}=\delta \bar{T}_{(\ref{heavy9}),R}
= C_F \frac{C_A}{2}\left[\frac{1+\left(1-\frac{y-\lambda_1^-}{1+\lambda_1^-} \right)\left(1-y\right)}{1+(1-y)^2}\left( \frac{y-\frac{\lambda_1^-}{2}}{y - \lambda_1^-}\right)\frac{\mathbf{l}_{\perp}\cdot \left(\mathbf{l}_{\perp} - \frac{y-\lambda_1^-}{1+\lambda_1^-}\mathbf{k}_{1 \perp}\right)}{\left(\mathbf{l}_{\perp} - \frac{y-\lambda_1^-}{1+\lambda_1^-}\mathbf{k}_{1 \perp} \right)^2} \right ].
\end{eqnarray}

\begin{figure}[thb]
\centering
\includegraphics[width=0.99\linewidth]{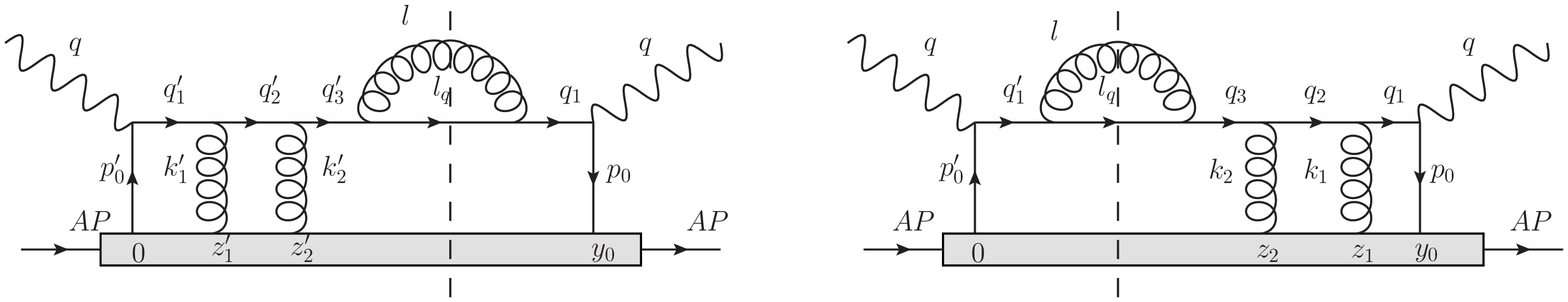}
 \caption{Two non-central-cut diagrams: two rescatterings in the amplitude and zero in the complex conjugate (or vice versa).}
 \label{heavy10}
\end{figure}

The phase factor for Figure~\ref{heavy10} reads:
\begin{eqnarray}
&&S_{(\ref{heavy10}),L}\,\,=\,\,-\frac{1}{2}e^{- i x_{D1} p^+ \delta z_1^-}(e^{i x_L p^+ Z_1^-}),
\nonumber\\
&&S_{(\ref{heavy10}),R}\,\,=\,\,-\frac{1}{2} e^{i x_L p^+ y_0^-}e^{ i x_{D1} p^+ \delta z_1^-}(e^{-i x_L p^+ Z_1^-}),
\nonumber\\
&&S_{(\ref{heavy10})}\,\,=\,\,S_{(\ref{heavy10}),L}\,\,+\,\,S_{(\ref{heavy10}),R}\,\,\approx\,\,-\cos(x_L p^+ Z_1^-).
\end{eqnarray}
The matrix element for Figure~\ref{heavy10} reads:
\begin{eqnarray}
\delta \bar{T}_{(\ref{heavy10}),L}=\delta \bar{T}_{(\ref{heavy10}),R}
=C_F^2.
\end{eqnarray}

\begin{figure}[thb]
\centering
\includegraphics[width=0.99\linewidth]{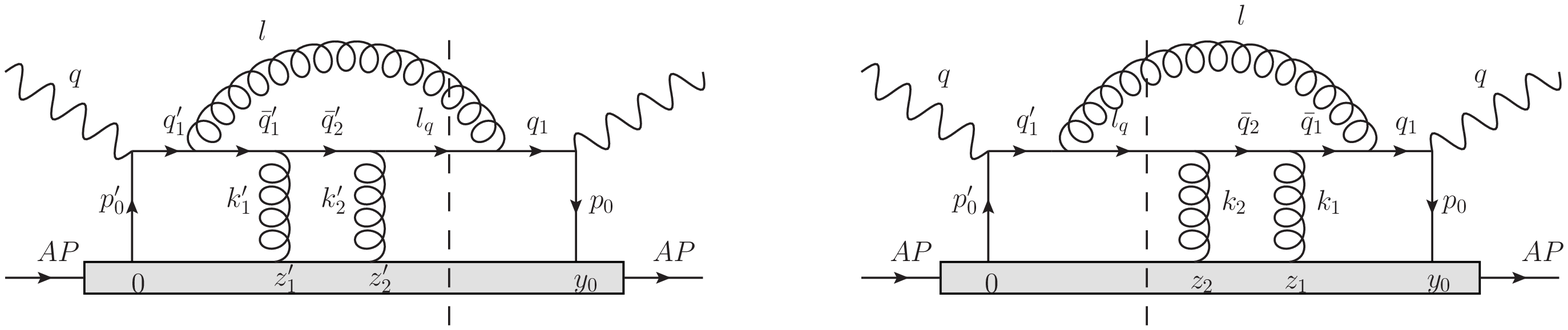}
 \caption{Two non-central-cut diagrams: two rescatterings in the amplitude and zero in the complex conjugate (or vice versa).}
 \label{heavy11}
\end{figure}

The phase factor for Figure~\ref{heavy11} reads:
\begin{eqnarray}
&& S_{(\ref{heavy11}),L}\,\,=\,\,-\frac{1}{2}e^{- i (y x_L-\eta_{D1}) p^+ \delta z_1^-}(1-e^{i x_L p^+ Z_1^-}),
\nonumber\\
&& S_{(\ref{heavy11}),R}\,\,=\,\,-\frac{1}{2}e^{i x_L p^+ y_0^-}e^{-i x_{D0} p^+ y_0^-}e^{i (y x_L-\eta_{D1}) p^+ \delta z_1^-}(1-e^{-i x_L p^+ Z_1^-}),
\nonumber\\
&& S_{(\ref{heavy11})}\,\,=\,\, S_{(\ref{heavy11}),L}\,\,+\,\, S_{(\ref{heavy11}),R}\,\,\approx\,\, \cos(x_L p^+ Z_1^-)-1.
\end{eqnarray}
The matrix element for Figure~\ref{heavy11} reads:
\begin{eqnarray}
\delta \bar{T}_{(\ref{heavy11}),L}=\delta\bar{ T}_{(\ref{heavy11}),R}
=C_F^2.
\end{eqnarray}

\begin{figure}[thb]
\centering
\includegraphics[width=0.99\linewidth]{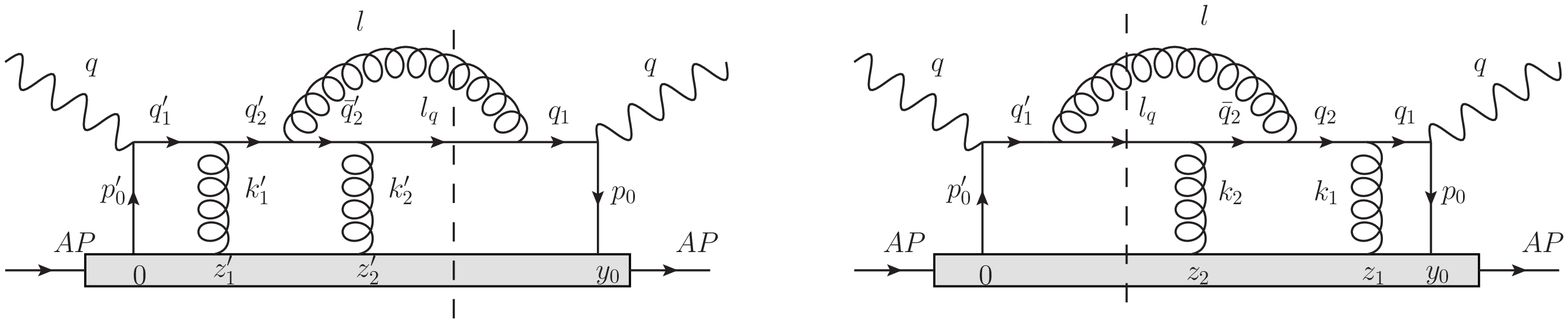}
 \caption{Two non-central-cut diagrams: two rescatterings in the amplitude and zero in the complex conjugate (or vice versa).}
 \label{heavy12}
\end{figure}

The phase factor for Figure~\ref{heavy12} reads:
\begin{eqnarray}
&& S_{(\ref{heavy12}),L}\,\,=\,\,-\frac{1}{2}e^{i x_L p^+ Z_1^-}\left(e^{i ((1-2y)x_L+2\eta_{D1}) p^+ \frac{1}{2}\delta z_1^-} - e^{-i (x_L  - 2 x_{D1}) p^+ \frac{1}{2}\delta z_1^-}\right),
\nonumber\\
&& S_{(\ref{heavy12}),R}\,\,=\,\,-\frac{1}{2}e^{i x_L p^+ y_0^-}e^{-i x_{D0} p^+ y_0^-}e^{-i x_L p^+ Z_1^-}\left(e^{-i ((1-2y)x_L+2\eta_{D1}) p^+ \frac{1}{2}\delta z_1^-} - e^{i (x_L  - 2 x_{D1}) p^+ \frac{1}{2}\delta z_1^-}\right),
\nonumber\\
&& S_{(\ref{heavy12})}\,\,=\,\,S_{(\ref{heavy12}),L}\,\,+\,\,S_{(\ref{heavy12}),R}\,\,\approx\,\, 0.
\end{eqnarray}
The matrix element for Figure~\ref{heavy12} reads:
\begin{eqnarray}
\delta \bar{T}_{(\ref{heavy12}),L}=\delta \bar{T}_{(\ref{heavy12}),R}
=C_F\left(C_F-\frac{C_A}{2}\right)\frac{1+(1-y)\left(1-\frac{y}{1+\lambda_1^-}\right)}{1+(1-y)^2}\left[\frac{\mathbf{l}_{\perp} \cdot \left(\mathbf{l}_{\perp} - \frac{y}{1+\lambda_1^-} \mathbf{k}_{1\perp}\right)}{ \left(\mathbf{l}_{\perp} - \frac{y}{1+\lambda_1^-} \mathbf{k}_{1\perp}\right)^2}\right].
\end{eqnarray}

\end{widetext}
%

\bibliographystyle{plain}
\bibliographystyle{h-physrev5}
\bibliography{refs_GYQ}
\end{document}